%% file: qrbst_avg_arXiv.tex
\documentclass[twocolumn,aps,prx,longbibliography,superscriptaddress]{revtex4-1}
\usepackage{float}
\usepackage{setspace}

\input{texdefs}

\input{texdefs1}

\usepackage[colorlinks=true,
            urlcolor=blue,
            citecolor=red,
            pdfstartview=FitH,
            pdfpagemode=None]{hyperref}

\newcommand{\black}[1]{\textcolor{black}{#1}}

\usepackage{times}
\usepackage{graphicx}
\usepackage{amssymb}
\usepackage{amsmath}
\usepackage{amsfonts}
\usepackage[dvips]{epsfig}

\usepackage{bbm}

\newcommand{\makethm}[3]{
\vspace{-1ex}
  \begin{thm}[#1]\label{thm:#2}
    #3
  \end{thm}
}

\usepackage[space]{grffile}

\newcommand{\figs}{figs}

\newcommand{\ddrefs}{\cite{Viola:98,ViolaKL:99,Uhrig:07,KhodjastehLidar:07,DDPRA:2011,XiaGotzLidar:2011}}
\newcommand{\ffrefs}{\cite{GreenETAL:2013,Soare2014,Kaby:2014,Paz-Silva2014,QCTRL:2021,HaasHamEngr:2019,Chalermpusitarak2021,Cerfontaine2021,Yang:2022}}

\newcommand{\ddffrefs}{\cite{Viola:98,ViolaKL:99,Uhrig:07,KhodjastehLidar:07,DDPRA:2011,XiaGotzLidar:2011,GreenETAL:2013,Soare2014,Kaby:2014,Paz-Silva2014,QCTRL:2021,HaasHamEngr:2019,Chalermpusitarak2021,Cerfontaine2021,Yang:2022}}

\newcommand{\toprefs}{\cite{RabitzHR:04,Ho.PRA.79.013422.2009,BeltraniETAL:2011,MooreR:2012,HockerETAL:2014,KosutArenzRabitz:2019}}

\newcommand{\nucnorm}[1]{\norm{#1}_{\rm nuc}}

\renewcommand{\mc}{{mN}}
\newcommand{\ellt}{{\ell t}}
\newcommand{\Obj}{\Phi}
\newcommand{\Jrbst}{J_{\rm rbst}}
\newcommand{\Fquad}{F_{\rm quad}}
\newcommand{\Jquad}{J_{\rm quad}}
\newcommand{\dv}{{\widetilde v}}

\newcommand{\cov}{{\rm cov}}

\newcommand{\tN}{{t=1:N}}
\newcommand{\tM}{{t=1:M}}

\renewcommand{\ns}{{n_S}}
\renewcommand{\nb}{{n_B}}
\renewcommand{\Is}{{I_\ns}}
\renewcommand{\Ib}{{I_\nb}}

\newcommand{\unc}{{\rm unc}}

\newcommand{\Vcalcon}{{\cal V}_{\rm ctrl}}
\newcommand{\Hcalunc}{{\cal H}_{\rm unc}}

\newcommand{\gamt}{\widetilde{\gam}}
\newcommand{\vb}{\bar{v}}
\newcommand{\vinc}{\widetilde{v}}
\renewcommand{\th}{\theta}

\renewcommand{\Hb}{{\bar{H}}}
\renewcommand{\Ht}{{\widetilde{H}}}

\newcommand{\Ebf}{{\bf E}}

\newcommand{\B}{{B}}
\newcommand{\Gavg}{G_{\rm avg}}
\newcommand{\vecGavg}{\vec{G}_{\rm avg}}
\newcommand{\Gb}{{G_{\rm avg}}}
\newcommand{\D}{D}
\renewcommand{\trace}{{\bf tr}}
\renewcommand{\vecc}{{\rm vec}}

\newcommand{\Fnom}{F_{\rm nom}}
\newcommand{\Fwc}{F_{\rm wc}}

\renewcommand{\Ub}{{\bar{U}}}
\newcommand{\normsm}[1]{\|#1\|}
\newcommand{\normsmf}[1]{\|#1\|_{\rm fro}}
\renewcommand{\normf}[1]{\norm{#1}_{\rm fro}}
\newcommand{\beasep}[1]{\renewcommand{\arraystretch}{#1}\bea}

\begin{document}

\title{Robust Quantum Control: Analysis \& Synthesis via Averaging}

\author{Robert L. Kosut}
\affiliation{SC Solutions, Sunnyvale CA, 94085}
\affiliation{Department of Chemistry,
  Princeton University, Princeton, NJ, 08544}

\author{Gaurav Bhole}
\affiliation{Department of Chemistry,
  Princeton University, Princeton, NJ, 08544}

\author{Herschel Rabitz}
\affiliation{Department of Chemistry,
  Princeton University, Princeton, NJ, 08544}



\date{\today}

\begin{abstract}

  \black{An approach is presented for robustness analysis and quantum
    (unitary) control synthesis based on the classic method of
    averaging. The result is a multicriterion optimization competing
    the nominal (uncertainty-free) fidelity with a well known
    robustness measure: the size of an interaction (error)
    Hamiltonian, essentially the first term in the Magnus expansion of
    an interaction unitary. Combining this with the fact that the
    topology of the control landscape at high fidelity is determined
    by the null space of the nominal fidelity Hessian, we arrive at a
    new two-stage algorithm. Once the nominal fidelity is sufficiently
    high, we approximate both the nominal fidelity and robustness
    measure as quadratics in the control increments. An optimal
    solution is obtained by solving a convex optimization for the
    control increments at each iteration to keep the nominal fidelity
    high and reduce the robustness measure.  Additionally, by
    separating fidelity from the robustness measure, more flexibility
    is available for uncertainty modeling.}

  \end{abstract}  

\maketitle

\flushbottom
\section{Introduction}

For a quantum computer that is \emph{ideally} composed of a temporal
sequence of unitary logic gates, robustness means that at specified
times, despite imperfections or perturbations, each \emph{actual} gate
operation is sufficiently close to its ideal for the intended purpose,
\ie, errors are all below a threshold value which allows for a quantum
computation.

In this paper, the standard state transformation used in the
\emph{method of averaging} \cite{Hale:80} is modified for quantum
systems resulting in a robustness framework useful for both analysis
and control synthesis. In particular, what arises from the theory is a
\emph{multicriterion control objective}: maximize the fidelity
corresponding to the control dependent nominal (perturbation-free)
unitary \emph{and} minimize a term (or terms) dependent on the the
characteristics of the uncertainty model.  Specifically, robustness is
measured by the size of the first term in the Magnus expansion of a
unitary evolving from a Hamiltonian dependent on the interaction of
the nominal unitary and the Hamiltonian uncertainty. Suppressing this
term is an underlying condition in the theoretical foundations of both
dynamical decoupling \ddrefs\ and optimization methods for general
noise suppression using filter functions \ffrefs.  In a similar way,
averaging theory establishes that if this term is made sufficiently
small, and the nominal fidelity is maximized, then the effect of
uncertain errors, which need not necessarily be small, can be
significantly alleviated.

In the approach presented here, the inherent multicriterion control
objective is intentionally maintained because of a special quantum
control landscape property: when the nominal fidelity is near its
maximum and the gradient is near zero, a large null space remains
specified by the Hessian \toprefs.  This freedom, which serves as a
space to project the system disturbances, also motivates a new
two-stage algorithm. First, only the nominal fidelity is maximized
until a high fidelity threshold is crossed. Then the optimization
minimizes the robustness measure while keeping the nominal fidelity
above the threshold.  The uncertainty-free optimization in the first
stage, assuming sufficient control resources, can be accomplished by
any iterative ascent method. For the second stage, we show that by
approximating both the nominal fidelity and the robustness measure as
quadratic functions in the control increments (via gradients and
Hessians), a robust control can be found by solving a convex
optimization problem at each iteration.

A second consequence of separating the robustness measure from the
nominal fidelity is that additional flexibility is provided for
\emph{uncertainty modeling}, \ie, incorporating more specific
characteristics of error sources in the robustness measure.  For
example, some of the errors delineated here seem naturally accounted
for by time-domain uncertainty modeling.  When the uncertainty, or
perturbation, can also be represented by a power spectral density, or
has distinct spectral lines, then the associated time-domain
robustness measure can be expressed equivalently in the frequency
domain using filter functions in a manner similar to that in \ffrefs.

The paper is organized as follows: \S\refsec{qsys} describes the basic
quantum system, \S\refsec{analysis} presents the averaging theory,
\S\refsec{synthesis} describes the multicriterion optimization problem
and the two-stage algorithm, \S\refsec{hcalunc} delineates some common
uncertainty models and associated interaction representations
(summarized in Table \ref{tabunc}), \S\refsec{numex} provides selected
representative numerical examples, and \S\refsec{sum} makes a few
concluding remarks. Appendices contain the averaging theory proof, a
detailed description of the second stage convex optimization, and some
additional uncertainty models and numerical examples.

\textbf{Notation} Continuous-time functions such as a unitary are
denoted by $U(t)$ over the time interval $t\in[0,T]$. For
discrete-time representations we use $t$ as a time index $\tN$ where
$U_t$ is the value of $U(t)$ at the end of each of $N$ uniformly
spaced intervals of width $T/N$. $I_n$ is the $n\times n$ identity
matrix.  We use a few norms: for an $n$-vector $x$, $\norm{x}_2$ is
the two-norm, for an $n\times m$ matrix $X$, $\normsm{X}_2$ is the
maximum singular value of the matrix, $\normsmf{X}$ is the Frobenius
matrix norm (\ie, the two-norm of the singular values),
$\normsm{X}_{\rm nuc}$ is the nuclear norm, the sum of the singular
values, and $\norm{X}_\infty$ is the max-row-sum.  We sometimes write
$\normsm{\cdot}$ with no subscript, usually meaning the norm will be
defined in context. For an $n\times m$ matrix $A$ we denote by $\vec
A$, the $nm\times 1$ vector formed by stacking the columns of $A$ from
left to right.

\section{Quantum system}
\label{sec:qsys}

The quantum systems addressed are mainly focused on \emph{closed}
systems. Our framework for robust control can easily include open
systems which are briefly discussed in \S\refsec{open}.  The closed
$n$-dimensional quantum system unitary evolution $U(t)$ over
$t\in[0,T]$ is obtained from,
\beq[eq:U]
\beasep{1.25}{ccl}
i\dot U(t) &=& \left(\Hb(t)+\Ht(t)\right)U(t),\
U(0)=I_n
\\
\Hb(t) &=& H_0 + H_c(t) 
\eea
\eeq
The Hamiltonian consists of two terms: the first, $\Hb(t)$, is the
\emph{nominal}, a known unperturbed term with a (usually) constant
drift term, $H_0$, and a time-varying control term, $H_c(t)$, which
depends on $m$ control variables $v_1(t),\ldots,v_m(t)$. In many cases
$H_c(t)$ is linear in the control variables, \ie, $H_c(t)=\sum_{j=1}^m
v_j(t)H_j$. The corresponding nominal (unperturbed) unitary, denoted
by $\Ub(t)$, evolves from,
\beq[eq:Unom]
i\dot \Ub(t) = \Hb(t)\Ub(t),\
\Ub(0)=I_n
\eeq
The nominal fidelity to achieve a unitary target $W$ is,
\beq[eq:Fnom]
\Fnom = |\trace(W^\dag\Ub(T)/n)|^2 \in[0,1]  
\eeq
The second term, $\Ht(t)$, is an uncertain \emph{perturbation} such
that,
\beq[eq:Hcalunc]
\left\{\Ht(t),t\in[0,T]\right\}\in\Hcalunc
\eeq
where the characteristics of $\Hcalunc$ are known, \eg, norm bounds
and/or probability distributions on Hamiltonian parameters or parts
thereof.
To assess the effect of a specified Hamiltonian perturbation on the
nominal system, it is natural to compare the final-time perturbed
system unitary $U(T)$ evolving from \refeq{U} with the target unitary
$W$. One such performance measure is the worst-case fidelity,
\beq[eq:Fwc]
\Fwc = \min_{\Ht\in\Hcalunc}|\trace(W^\dag U(T)/n)|^2,
\eeq
As described in \cite{KosutGB:13}, $\Fwc$ can be approximated by
sampling over the uncertainty set, and for small errors a convex
approximation along with a complementary optimization can provide more
confidence towards having found the worst-case fidelity. Similar
results hold for the average-case fidelity. However, as more error
types and sources are discovered, or others are considered, sampling
or convex approximations can get computationally burdensome,
particularly if multiple error sources have to be simultaneously
evaluated.

\section{Analysis of robustness}
\label{sec:analysis}



\black{To analyze the effect of perturbations we view the system
  evolution via
the \emph{interaction} unitary between the nominal and perturbed
unitaries \ddffrefs.}
Specifically, let $U(t)=\Ub(t)R(t)$ where $R(t)$ is the interaction
unitary which evolves from,
\beq[eq:Rt]
\beasep{1.25}{rcl}
i\dot R(t) &=& G(t)R(t),\ R(0)=I_n
\\
G(t) &=& \Ub(t)^\dag\Ht(t)\Ub(t)
\eea
\eeq
with $G(t)$ referred to as the \emph{interaction Hamiltonian.}  The
worst-case fidelity \refeq{Fwc} is then equivalently given by,
\beq[eq:FwcR]
\Fwc = \min_{\Ht\in\Hcalunc}|\trace(W^\dag\Ub(T)R(T)/n)|^2
\eeq
%
Deviations from the nominal are now captured by the deviation of
$R(T)$ from identity, \ie, if $R(T)\approx I_n$ then
$\Fwc\approx\Fnom$. This approximation can be made precise by
utilizing the standard method of averaging \cite{Hale:80} which shows
a means for significantly reducing the size of $\normsm{R(T)-I_n}$
by controlling the \emph{nominal trajectory} $\Ub(t),t\in[0,T]$.
The following result is established in Appendix \S~\refsec{avg
  analysis}.

%
%

\makethm{Averaging Analysis}{avg}{
For the unitary system \refeq{U}-\refeq{Unom}, 
denote by $\Gavg$ the {time-averaged Hamiltonian},
\beq[eq:Gavg]
\Gavg=\frac{1}{T}\int_0^T\Ub(t)^\dag\Ht(t)\Ub(t)dt 
\eeq
Assume that amongst the properties of the uncertainty set
\refeq{Hcalunc} is the bound,
%
$\normsm{\Ht(t)} \leq \del$ for a specified norm. 
%
Under these conditions,
\beq[eq:rbst]
\beasep{2}{ll}
\mbox{if}
&\normsm{\Gavg}\leq\gamb~\del
\quad\mbox{and}\quad
\normsm{G(t)-\Gavg}\leq\gamt~\del,
\\
\mbox{then} & \mbox{as $\Fnom\to 1$,}
\\
&\Fwc \to
\Fnom - \left(\Ocal\{\gamb(\del T)\} + \Ocal\{\gamt(\del T)^2\}\right)^2
\eea
\eeq

\mbox{}
}  


\noindent
Theorem~\ref{thm:avg} shows that if the nominal performance is
acceptable, \ie, $\Fnom\approx 1$, \emph{and} the mean ($\gamb$) and
deviation ($\gamt$) of the \emph{time-averaged Hamiltonian}
(effectively the first term in the Magnus expansion of $R(T)$, the
interaction unitary) are sufficiently small, then significant robustness
about the nominal accrues. And of note, $\del T$ need not be
small. However, if only the mean $\gamb$ can be made very small, then
at most only higher order errors are present.
The worst-case fidelity summarized in \refeq{rbst} follows
from an explicit bound as part of the derivation provided in Appendix
\S~\refsec{avg analysis}.

The analysis result above is readily extended to systems with multiple
\emph{additive} error sources and associated time-averaged
Hamiltonians, that is,
%
\beq[eq:Lpert]
\beasep{1.5}{rcl}
\Ht(t) &=& \sum_{\ell=1}^L\Ht_\ell(t),\quad
\Ht_\ell(t)\in\Hcalunc^\ell
\\
\Gavg^\ell &=& \frac{1}{T}\int_0^T\Ub(t)^\dag\Ht_\ell(t)\Ub(t)dt
\eea
\eeq
Large values of any $\Gavg^\ell$ indicates a source of potentially
disruptive errors. Thus each of the time-averaged Hamiltonians can be
used to assess the impact on the nominal design of any anticipated or
suspected errors \emph{without having to run full simulations} and
without resorting to a possibly computationally burdensome worst-case
error evaluation \refeq{Fwc}. 

These properties of the time-averaged Hamiltonian show in general how
the \emph{trajectory} of the nominal propagator can establish
robustness.  As there are many pathways via control that achieve the
same fidelity level \toprefs.  there is expected to be considerable
freedom in choosing a pathway that simultaneously achieves a high
fidelity nominal response \emph{and} minimizes the size of each of the
time-averaged Hamiltonians $\Gavg^\ell$. The freedom discussed arises
from the ability to roam over the null space at the top of the
fidelity landscape.
%

\section{Robust control synthesis}
\label{sec:synthesis}

\subsection{Synthesis criteria}

Theorem~\ref{thm:avg} also provides design criteria to \emph{synthesize}
a robust control.  Specifically, the \emph{nominal} propagator
$\Ub(t),t\in[0,T]$ should have the following features:
\ben\setlength{\itemsep}{0.15in}
\item the final time nominal unitary is very close to the target $W$,
  \ie, the nominal fidelity $\Fnom \approx 1$ \refeq{Fnom}.

\item each time-averaged Hamiltonian perturbation \refeq{Lpert} is
  small: $\Gavg^\ell\approx 0$.

\een
These two goals, which place simultaneous demands on the \emph{nominal
  trajectory}, $\Ub(t),t\in[0,T]$, has been presented in various forms
in \ffrefs.

Here we provide another version formulated as a multicriterion
optimization problem.  The synthesis (or design) variables are the
controls $v_j(t),j=1:m,t\in[0,T]$ which are constrained to a set
$\Vcalcon$ reflecting any limitations imposed by the control
actuators, \eg, magnitude, sampling rate, bandwidth limitations from
actuator dynamics, linear combination of basis functions, \eg, for a
piece-wise-constant constraint the basis functions are pulses, \etc.


\subsection{Multicriterion optimization}

To satisfy the two synthesis criteria arising from
Theorem~\ref{thm:avg}, we form the following generic
\emph{multicriterion} optimization problem,
\beq[eq:Jgen]
\beasep{1,5}{ll}
\mbox{minimize}&
\Obj(v) = \left\{
1-\Fnom(v),\ \Jrbst(v)
\right\}
\\
\mbox{subject to}&
\Jrbst(v) =
\sum_\unc\del_\unc\norm{\Acal_\unc(v)S_\unc}_\unc
\\&
v\in\Vcalcon
\eea
\eeq
where $v$ is the vector of control variables, $\Jrbst(v)$ is a
robustness measure reflecting the magnitude of the time-averaged
Hamiltonians in \refeq{Lpert dis}, and the objective notation
$\{\cdot,\cdot\}$ is meant to denote the possible trade-off between
the two objectives \cite{BoydV:04}.

In a later section, where we describe some common types of
uncertainty, we show that all of these result in a robustness measure
$\Jrbst(v)$ which has the generic form shown. Specifically, the matrix
$\Acal_\unc(v)$ will be seen to capture the interaction between the
nominal unitary trajectory and the uncertainty, the matrix $S_\unc$,
the norm $\norm{\cdot}_\unc$ and associated error bound $\del_\unc$
all follow from the probabilistic or deterministic uncertainty
character.
With a single uncertainty, or several with near equal bounds, it is
not always necessary to include the bounding variables, in which case
they are effectively absorbed into the robustness measure.

\subsection{Two-stage algorithm}
\label{sec:twoalg}

As alluded to earlier, at the top of the control landscape where
$\Fnom(v)=1$, the gradient $\nabla_v\Fnom(v)=0$, there is potentially a
large null space \toprefs.
This \emph{roaming} freedom at the top of the landscape derives from
the structure of the Hessian of the nominal fidelity,
$\nabla_v^2\Fnom(v)$, which is negative semidefinite and low-rank, 
and this can only have a modest number of eigenvectors with negative
eigenvalues (\ie, these specify the portion of the control that
maximizes the fidelity measure). In turn, the remaining, often very
large null space, with associated zero eigenvalues, can serve as a
space to project the system disturbances. Ergo, one iterative path to
finding a solution to \refeq{Jgen} is to first maximize only the
nominal fidelity $\Fnom(v)$.  When this fidelity crosses a high
threshold, $f_0\approx 1$, then switch to minimizing the robustness
measure $\Jrbst(v)$ while keeping the fidelity above $f_0$.
Specifically, we utilize the following two stage algorithm:
\beq[eq:twostage]
\beasep{1.25}{l}
\mbox{{\bf Stage 1} maximize {\bf only}
  $\Fnom(v)$ {\bf until} $\Fnom(v) \geq f_0$}
\\
\mbox{{\bf Stage 2} minimize $\Jrbst(v)$ {\bf while} $\Fnom(v) \geq f_0$}
\eea
\eeq
The uncertainty-free optimization in Stage 1 can be done by many
different gradient ascent methods. For Stage-2, when $\Fnom(v)\geq
f_0$, we show in Appendix \S\refsec{comp}
that finding an optimal control increment $\widetilde v$ to solve
\refeq{twostage} can be recast \emph{exactly} as a convex optimization
problem. This approach differs from earlier works which used the
properties of the null space at high fidelity directly \toprefs.

In particular, during Stage-2, let fidelity and robustness at the
current control $v$ be denoted by $\Fnom$ and $\Jrbst$. For a change
in controls from $v$ to $v+\dv$, by approximating both the fidelity
and robustness changes by quadratic functions (gradients and Hessians)
in the control increments $\dv$, the next control increment can be
found by solving the following optimization problem:
%
\beq[eq:quad]
\beasep{1.5}{ll}
\mbox{minimize}&
\dv^T\dv+\alf(\Jquad(\dv)-\Jrbst)
\\
\mbox{subject to}&
\Fquad(\dv) \geq f_0
\eea
\eeq
where
\beq[eq:FJquad]
\beasep{1.5}{rl}
\Fquad(\dv) &= \Fnom+\nabla\Fnom^T\dv
+ \dv^T\nabla^2\Fnom\dv/2
\\
\Jquad(\dv) &= \Jrbst+\nabla\Jrbst^T\dv
+ \dv^T\nabla^2\Jrbst\dv/2
\eea
\eeq
The objective is a trade-off between a small control increment $\dv$
and a $\Jrbst$ decrease (robustness increase) with robustness weight
$\alf>0$. The constraint keeps the nominal fidelity higher than
$f_0\approx 1$. Although \refeq{quad} is not always a convex
optimization, using \cite[App.B]{BoydV:04} as shown in Appendix
\S\refsec{comp}, the increment $\dv$ can be found efficiently by
solving for the two variables $\gam,\lam\in\Rbf$ in the associated
(convex) dual optimization problem,
\beq[eq:dual0]
\beasep{1.25}{l}
\mbox{maximize}\quad \gam
\\
\mbox{subject to}\quad
\lam\geq 0
\\
\mat{I-\lam \nabla^2\Fnom+\alf\nabla^2\Jrbst
  &\alf \nabla\Jrbst-\lam\nabla\Fnom
\\
(\alf\nabla\Jrbst-\lam\nabla\Fnom)^T&-2\lam(\Fnom-f_0)-\gam} \geq 0
\eea
\eeq
From $\lam$ the control increment is given by,
\beq[eq:dv opt]
\dv = -\left(I-\lam\nabla^2\Fnom+\alf\nabla^2\Jrbst\right)^{-1}
(\alf\nabla\Jrbst-\lam\nabla\Fnom)
\eeq
In addition, if the threshold is maintained, \ie, $\Fquad(\dv)>f_0$,
then $\Jrbst$ can never
increase. Figures~\ref{fig:fig1const}-\ref{fig:fig1tv} show two basic
examples of the two-stage algorithm that exhibit the topological
property at high fidelity.

\section{Uncertainty Modeling}
\label{sec:hcalunc}


In this section several representative Hamiltonian uncertainty sets
\refeq{Hcalunc} are presented along with their corresponding
robustness measures $\Jrbst$ \refeq{Jgen}.  The uncertainty is
represented by unknown but bounded Hamiltonian parameters that can be
constant or time varying, and either deterministic or
probabilistic. The models are canonical, their main purpose being to
capture uncertainty obtained from the underlying physics supported by
data. If they are too conservative in relation to the control
resources, then robustness will be compromised.

A few of the uncertainty models described in this section are
displayed in Table-\ref{tabunc}. Clearly there are many combinations
and variations of these elementary uncertainty sets that can be
formulated. Selected numerical examples are presented in
\S\ref{sec:numex}. Additional uncertainty models are presented in
Appendix \S\refsec{more uncmod} including multiplicative control noise,
actuator dynamic uncertainty, cross-coupling, open bipartite systems, and
Lindblad uncertainty. A few additional numerical examples are in
Appendix \S\refsec{more numex}.


\begin{table*}[t]
  \setlength{\tabcolsep}{9pt} 
  \renewcommand{\arraystretch}{1.25} 
  \btab{|c|c|c|}
  \hline
  Uncertainty Type
  & \btab{c}Uncertainty Set\\$\Ht_t\in\Hcalunc$\etab
  & \btab{c}Robustness Measure\\
  $\Jrbst = \del_\mu\norm{\Acal S}_\mu$
  \etab
    \\
  \hline\hline
  \btab{c}Constant parameters\\(magnitude bounded)
  \\\mbox{(see \S\refsec{unc const})}
  \etab
  &
  $\bea{c}
  \Ht_t = \sum_{\ell=1}^L\th_\ell B_{\ell t}
  \\
  \mbox{$B_{\ell t}$ known}
  \\
  \normsm{\vec\th}_\mu \leq \del_\mu
  \eea
  $
  & 
  $\bea{c}
  \Acal = \mat{\vec A_1&\cdots&\vec A_L}\ (n^2\times L)
  \\
  A_\ell = (1/M)\sum_{t=1}^M\Ub_t^\dag B_{\ell t}\Ub_t,\ \ell=1:L
    \\
  \bea{lll}
  \normsm{\vec\th}_\infty\leq\del_\infty&\Rightarrow&\mu=\infty,\ S=I_L
  \\
  \normsm{\vec\th}_2\leq\del_2&\Rightarrow&\mu=2,\ S=I_L
  \\
  \cov~\vec\th = C &\Rightarrow&\mu={\rm fro},\ S=\sqrt{C}
  \eea
  \eea
  $
  \\
  \hline
  \btab{c}Constant Hamiltonian\\(energy-bounded)
  \\\mbox{(see \S\refsec{energy bnd})}
  \etab
  & $\bea{c}
  \mbox{$\Ht$ constant,\ $\normsmf{\Ht}\leq\del$}
  \\\mbox{equivalent to:}\\
  \Ht=\sum_{i=1}^{n^2}\th_i\Gam_i,\ \normsm{\vec\th}_2\leq\del
  \\
  \mbox{
    $\mat{\vec\Gam_1&\cdots&\vec\Gam_{n^2}}$,\ $(n^2\times n^2)$ unitary
    }
  \eea$
  & $\bea{c}
  \Acal = \mat{\vec A_1&\cdots&\vec A_{n^2}}
  \quad(n^2\times n^2)
  \\
  A_i = \frac{1}{M}\sum_{t=1}^M\Ub_t^\dag\Gam_i\Ub_t,\
  i=1:n^2
  \\
  S=I_{n^2},\ \mu=2
  \eea
  $
  \\\hline
  \btab{c}Time-varying parameter\\(band limited, stationary)
  \\
  \mbox{(see \S\refsec{Hpartv})} 
  \etab
  & $\bea{c}
  \Ht_t = \th_t B_t
  \\\mbox{band-limited constraint:}\\
  \vec\th = \del\left(\Kcal\vec{w}/\normsmf{\Kcal}\right),\
  \cov~\vec w = I_M
  \\
  \mbox{$\Kcal$ is Toeplitz noise filter matrix}
  \eea$
  &
  $\bea{c}
  \Acal = \mat{\vec A_1&\cdots&\vec A_M}\ (n^2\times M)
  \\
  A_t = \Ub_t^\dag\B_t\Ub_t/M,\ t=1:M
  \\
  S = \Kcal/\normsmf{\Kcal},\ \mu={\rm fro}
  \eea
  $
  \\\hline
  \etab
    \caption{Table summarizing three common classes of Hamiltonian
      uncertainties and associated robustness measures \refeq{Jgen} as
      described in \S\refsec{hcalunc}. \\{\bf Key:} system dimension
      is $n$, PWC control sample rate is $T/N$; averaging/simulation
      sample rate is $T/M$ with $M/N$ an integer. }
  \label{tabunc}
\end{table*}


\subsection{Piece-wise-constant (PWC) control}
\label{sec:pwc}

To make the uncertainty modeling procedures precise, we start with the
minimal restriction that the ideal control actuator produces $N$
perfect pulses of width $T/N$. This motivates switching to a
discrete-time picture, where now $\tN$ denotes these $N$ uniformly
spaced intervals. Equivalently, each of the $m$ controls in \refeq{U}
are piece-wise-constant (PWC) with magnitudes $v_{tj},j=1:m,\tN$;
these $mN$ magnitudes are the design variables. (In general the PWC
controls could be phases, frequencies or weights in some basis.)
%

The nominal propagator evolution is then determined by a
product of $N$ unitaries, expressed recursively for $\tN$ as,
%
\beq[eq:Unom dis]
\beasep{1.25}{rcl}
\Ub_t &=& e^{-i(T/N)\Hb_t}\Ub_{t-1},\ \Ub_0=I_n
\\
\Hb_t &=& H_0 + H_c(v_t), 
\quad v_t=\{v_{tj},j=1:m\}
\eea
\eeq
The discrete-time version of each of the $L$ time-averaged
Hamiltonians in \refeq{Lpert} is obtained at an appropriate sampling
rate for the uncertainty characteristics. This rate is not necessarily
the same as the control PWC rate of $T/N$. Let the sampling rate for
averaging (and simulation) be $T/M$ with $M\geq N$ and for consistency
$M/N$ is an integer. Then,
\beq[eq:Lpert dis]
\beasep{1.5}{rcl}
\Ht_t &=& \sum_{\ell=1}^L\Ht_\ellt,\quad 
\Ht_\ellt\in\Hcalunc^\ell
\\
\Gavg^\ell &=& \frac{1}{M}\sum_{t=1}^M\Ub_t^\dag\Ht_\ellt\Ub_t
\eea
\eeq
For ease of notation we use $t$ as the sample time index if it is
clear in the context which rate applies,, \ie, $T/N$ for PWC control
\refeq{Unom dis} and $T/M$ for uncertainty and averaging \refeq{Lpert
  dis}.


\subsection{Uncertain constant Hamiltonian parameter}
\label{sec:unc const}

A common form of an uncertain perturbation is one composed of the sum
of products of unknown but bounded \emph{constant} parameters and
known (possibly time-varying) Hamiltonians. The resulting Hamiltonian
uncertainty set is,
\beq[eq:Hcal th]
\Hcalunc = \left\{
\Ht_t=\sum_{\ell=1}^L\th_\ell\B_\ellt\
\left|\
\bea{l}
{\vec\th} = \mat{\th_1&\cdots&\th_L}^T \in\Theta
\\
\norm{\B_\ellt}\leq 1,\ t=1:M
\eea
\right.
\right\}
\eeq
The corresponding time-averaged Hamiltonian is,
\beq[eq:Gavgpar]
\Gavg = \frac{1}{M}\sum_{t=1}^M\sum_{\ell=1}^L
\th_\ell\Ub_t^\dag\B_\ellt\Ub_t 
\eeq
%
\subsection{Bounded parameters}
\label{sec:mag bnd}

To model the effect of uncertain but bounded Hamiltonian parameters,
assume that the $L\times 1$ vector of parameters $\vec\th$ belongs to
one of the following three uncertainty sets:
\beq[eq:parset]
\beasep{1.5}{cl}
\mbox{peak}&
\Theta_\infty = \{ \normsm{\vec\th}_\infty\leq\del_\infty\}
\\
\mbox{energy}&
\Theta_2 = \{\normsm{\vec\th}_2\leq\del_2\}
\\
\mbox{covariance}&
\Theta_{\rm prob} = \left\{
\bea{c}
\mbox{$\vec\th$ zero-mean}
\\
\cov~\vec\th = C\geq 0
\eea
\right\}
\eea
\eeq
One way to see the effect of the uncertain parameters is to stack
\emph{all} $n^2$ elements of $\Gavg$ in \refeq{Gavg} into the vector
$\vecGavg$ resulting in,
\beq[eq:vecGavg par]
\vecGavg = \Acal~\vec\th\quad
\left\{
\beasep{1.5}{l}
\Acal= \mat{\vec A_1&\cdots&\vec A_L} \in \Cbf^{n^2\times L}
\\
A_\ell = (1/M)\sum_{t=1}^M\Ub_t^\dag\B_\ellt\Ub_t \in\Cbf^{n\times n}
  \eea
  \right.
\eeq
Since $\vecGavg$ in linear in $\vec\th$, the robustness measures
corresponding to the sets in \refeq{parset} follow directly from the
induced norm properties, \ie,
\beq[eq:Jparset]
\beasep{1.5}{lcl}
\vec\th\in\Theta_\infty
&\Rightarrow&
\Jrbst = \del_\infty\normsm{\Acal}_\infty
= \del_\infty{\ds\max_{i=1:n^2}}\sum_{\ell=1}^L |\vec{A}_{\ell i}|
\\
\vec\th\in\Theta_2
&\Rightarrow&
\Jrbst = \del_2\normsm{\Acal}_2
\\
\vec\th\in\Theta_{\rm prob}
&\Rightarrow&
\Jrbst = \trace(\cov~{\Gavg})
=\normsm{\Acal\sqrt{C}}_{\rm fro}^2
\eea
\eeq
These robustness measures reflect individual terms in \refeq{Jgen},
the $\Acal$-matrix being the same for all three cases above.  For the
two deterministic sets ($\Theta_\infty,\Theta_2$) the effective
$S$-matrix is simply $S=I_L$ with $\mu=\infty$ and $\mu=2$,
respectively. For the probabilistic set ($\Theta_{\rm prob}$),
$S=\sqrt{C}$ with $\mu={\rm fro}$. Clearly more nuanced uncertainty
configurations are possible, \eg, mixing matrix and vector norms such
as $\Jrbst =\max_{\th\in\Theta_2}\normsm{\Gavg}_\infty$, and so on. In
addition, in a sum of measures, each measure could differ, or for
groups of parameters each might have different types of bounds.

\subsection{Bias and drift}
\label{sec:r2r}

Although $\vec\th$ in the instances described is a bounded uncertain
constant, the specific value within the bound can vary over each run
time interval $t\in[0,T]$. Thus the uncertainty set \refeq{Hcal th} is
describing parameters which can change abruptly from run-to-run. If,
however, $\vec\th_t$ is slowly drifting in each run, and is memoryless
from run-to-run, then for each of $L$ parameters a more suitable model
for $t=1:M$ is,
\beq[eq:th r2r]
\th_{\ell,t} = h_t^Tc_\ell,\quad
h_t= \frac{1}{M-1}\mat{M-t\\t-1},
\quad
c_\ell=\mat{a_\ell\\b_\ell}
\eeq
In each run $a_\ell$ is the initial value and $b_\ell$ is the
final value. Assuming that for some norm each $c_\ell$ is
independently bounded by $\normsm{c_\ell}\leq\del_\ell$, then
following \refeq{vecGavg par}-\refeq{Jparset}, the robustness measure
is,
\beq[eq:Jr2r]
\Jrbst = \sum_{\ell=1}^L\del_\ell\normsm{\Acal_\ell}
\quad
\left\{
\beasep{1.5}{l}
\Acal_\ell = \sum_{t=1}^M\vec A_{\ell,t}h_t^T \in\Cbf^{n^2\times 2}
\\
A_{\ell,t} = \Ub_t^\dag\B_{\ell,t}\Ub_t
\eea
\right.
\eeq
%

\subsection{Uncertain constant energy-bounded Hamiltonian}
\label{sec:energy bnd}

An interesting example of a constant Hamiltonian perturbation is the
set of uncertain energy-bounded constant Hamiltonians,
\beq[eq:ham bnd1]
\Hcalunc=\{
\Ht\ \mbox{constant},\ \normsmf{\Ht}\leq\del
\}
\eeq
Clearly $\Ht$ is arbitrary other than subject to an ``energy-like'' bound.
Consequently, an equivalent uncertainty set is,
\beq[eq:ham bnd2]
\Hcalunc=\left\{
\Ht=\sum_{i=1}^{n^2}\th_i\Gam_i\
\left|
\beasep{1.25}{l}
\mat{\vec\Gam_1&\cdots&\vec\Gam_{n^2}}\in\Ucal(n^2)
  \\
  \normsm{\vec\th}_2\leq\del
  \eea
 \right.
\right\}
\eeq
where $\Gam_i,i=1:n^2$ is any orthonormal basis set for $n\times n$
Hermitian matrices, and where the vector $\vec\th\in\Rbf^{n^2}$,
bounded in the two-norm by $\del$, consists of all the constant
coefficients $\th_i\in\Rbf,i=1:n^2$. The equivalence follows from the
fact that $\normsmf{\Ht}=\normsm{\vecc(\Ht)}_2$ along with the induced
norm property of the vector two-norm. Thus for any orthonormal
(Hermitian) basis, the real coefficients are equivalent to all
possible entries in \emph{any constant} $\Ht$ such that
$\normsmf{\Ht}\leq\del$.
It follows that for an energy bounded parameter uncertainty the
appropriate robustness measure is of the same form of
\refeq{Jparset}, \ie,
\beq[eq:JHenergy]
\Jrbst = \del\norm{\Acal}_2\quad
\left\{
\beasep{1.5}{l}
\Acal = \mat{\vec A_{1}&\cdots&\vec A_{n^2}}
\\
A_{i} = (1/M)\sum_{t=1}^M\Ub_t^\dag\Gam_i\Ub_t
\eea
\right.
\eeq
where the $S$-matrix is simply $I_{n^2}$, the identity matrix. 


\subsection{Uncertain time-varying Hamiltonian parameter}
\label{sec:Hpartv}

Uncertain parameter time variation can arise in a number of forms
depending on the underlying physics. Here we pose a few generic
uncertainty models which capture the effects of limited-bandwidth and
stationary noise.

\subsubsection{Limited bandwidth}

Let $\th_t,\tM$ denote an uncertain bounded \emph{time-varying}
parameter associated with the \emph{single} perturbation
$\Ht_t=\th_t\B_t$ where $B_t$ is known and typically normalized so
that $\norm{B_t}\leq 1$.  To capture limited-bandwidth, assume that
$\th_t$ is the output of a ``filter'' driven by a bounded sequence
$w_t,\tN$, where the filter is well represented as a
linear-time-varying (LTV) system with causal (convolution) dynamics
\cite{Ljung:87},
\beq[eq:LTV]
\th_t = \sum_{\tau=1}^t K_{t\tau}w_\tau
\eeq
%
Since causality implies that $K_{t\tau}=0$ for $\tau>t$, the vector of
time-varying parameters can be expressed by 
$\vec\th=\Kcal\vec w$ with $\Kcal$ the $M\times M$
\emph{block-lower-triangular} matrix,
\beq[eq:BLT]
\Kcal=\mat{K_{11}&&&\\K_{22}&K_{21}&&\\
    \vdots&\ddots&\ddots&\\
    K_{MM}&\cdots&K_{M2}&K_{M1}}
\eeq
If the filter is linear-time-invariant (LTI) then \refeq{LTV} becomes
$\th_t=\sum_{\tau=1}^t K_{t-\tau}w_\tau$ from which is follows that
the matrix $\Kcal$ is lower-triangular Toeplitz, \ie,
\beq[eq:toep]
\Kcal=\mat{K_{1}&&&\\K_{2}&K_{1}&&\\
    \vdots&\ddots&\ddots&\\
    K_{M}&\cdots&K_{2}&K_{1}}
\eeq
For an LTI filter the first column above is the impulse response of
the filter dynamics.

\subsubsection{Stationary noise modeling}
\label{sec:noise}


For either LTV or LTI dynamics, assume further that $\vec w$, the
noise filter input, is a zero-mean stationary sequence with unit covariance
$\cov(\vec w)=I_M$.  Under these conditions, the Hamiltonian
uncertainty set is,
\beq[eq:Hcal tht]
\Hcalunc=\left\{
\Ht_t=\th_t\B_t\ 
\left|
\bea{l}
\vec\th=\Kcal\vec w,\ \vec w \mbox{ zero-mean}
\\
\cov(\vec w)=I_M
\eea
\right.
\right\}
\eeq
The corresponding $n^2\times 1$ vector of the time-averaged Hamiltonian is,
\beq[eq:acal]
\vecGavg = \Acal \Kcal\vec w
\quad
\left\{
\beasep{1.5}{l}
\Acal = \mat{\vec A_1&\cdots&\vec A_N}\ (n^2\times M)
\\
A_t = \Ub_t^\dag\B_t\Ub_t/M 
\eea
\right.
\eeq
Consequently, for a single error source the output magnitude level is
subsumed in the noise filter $\Kcal$. (Explicit bound shown in
Table-\ref{tabunc}.)
For multiple time-varying 
perturbations of the form $\Ht_\ellt = \th_\ellt\B_\ellt,\ell=1:L$
with independent time-varying parameters $\vec\th_\ell=\Kcal_\ell\vec
w_\ell$, the robustness measure is the sum,
\beq[eq:J thtv mult]
\Jrbst = \sum_{\ell=1}^L
\normf{\Acal_\ell  S_\ell}^2
  \quad
  \left\{
  \bea{l}
  \Acal_\ell=\mat{\vec A_{\ell 1}&\cdots&\vec A_{\ell M}}
  \\
  A_{\ell t} = \Ub_t^\dag B_{\ell t}\Ub_t
  \\
  S_\ell = \Kcal_\ell
  \eea
  \right.
  \eeq
with $(S_\ell,\Acal_\ell)$ correspondingly defined as previously per
the characteristics of each perturbation and where output power of
each time-varying parameter is contained in each $\Kcal_\ell$.  As
shown in Appendix \S\refsec{fltr func}, in this instance the
robustness measure can be formed in the frequency domain using a filter
function as in \ffrefs.

\section{Numerical examples}
\label{sec:numex}

The purpose of these numerical examples is to illustrate the averaging
theory in conjunction with the the two-stage optimization as outlined
in \refeq{twostage}-\refeq{dv opt} and further described in
Appendix \S\refsec{comp}. 
The examples all demonstrate ready access to the null space at the top
of the control landscape which is essential for robust control. The
focus throughout is (mostly) on a few basic single qubit systems.
Appendices \S\refsec{more uncmod} and \S\refsec{more numex} contain
additional uncertainty models and numerical examples.

\begin{figure*}[t]
  \centering
  \btab{ccc}
  Infidelity \& robustness vs. iterations
  &
  Evaluation of infidelity vs. uncertainty
  &
  Control pulses: nominal vs. optimal
  \\
  \btab{c}
  \includegraphics[width=0.33\textwidth]{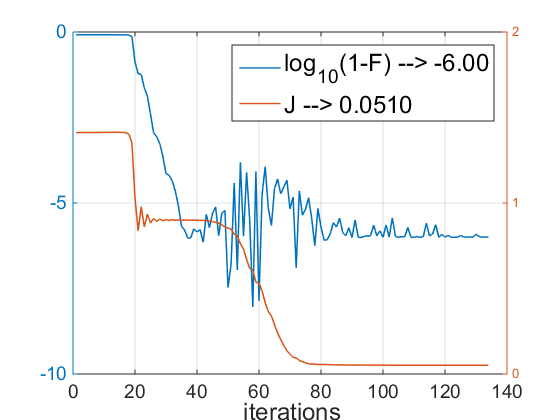}
  \\(a)
  \etab
  &
  \btab{c}
  \includegraphics[width=0.33\textwidth]{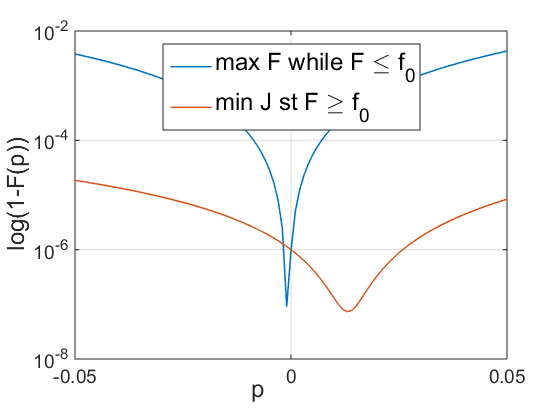}
  \\(b)
  \etab
  &
  \btab{c}
  \includegraphics[width=0.33\textwidth]{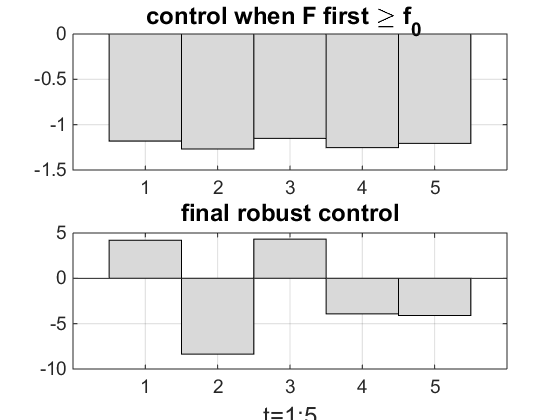}
  \\(c)
  \etab
  \etab
  \caption{{\bf Uncertain constant parameter}
    $H_t=v_t\sig_x+(1+\th)\sig_z$.}
  \label{fig:fig1const}
  \end{figure*}

  \begin{figure*}[t]
  \centering
  \btab{ccc}
  Infidelity \& robustness vs. iterations
  &
  Evaluation of infidelity vs. uncertainty
  &
  Control pulses: nominal vs. optimal
  \\
  \btab{c}
  \includegraphics[width=0.33\textwidth]{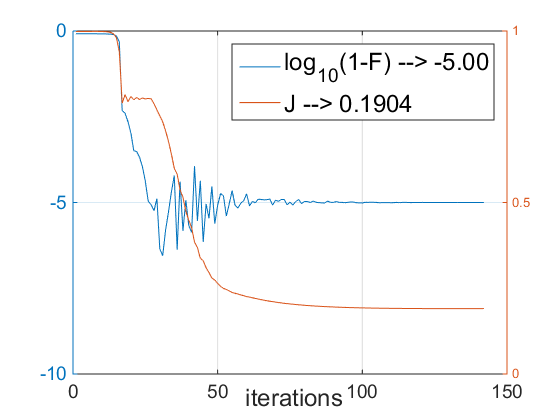}
  \\(a)
  \etab
  &
  \btab{c}
  \includegraphics[width=0.33\textwidth]{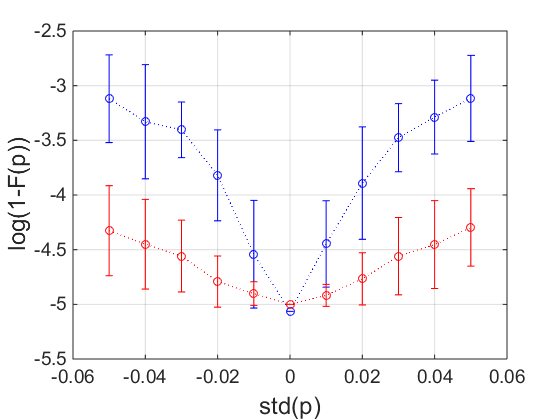}
  \\(b)
  \etab
  &
  \btab{c}
  \includegraphics[width=0.33\textwidth]{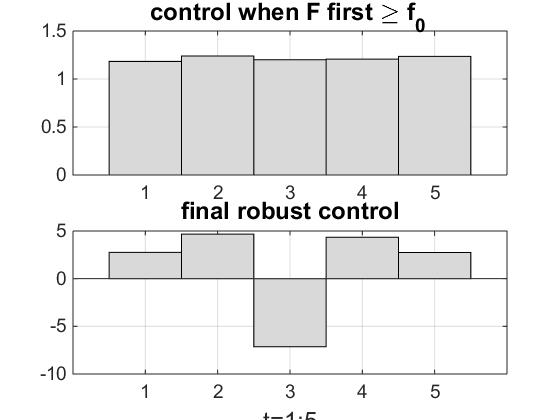}
  \\(c)
  \etab
  \etab
  \caption{{\bf Uncertain
      time-varying parameter} $H_t=v_t\sig_x+(1+\th_t)\sig_z$.}
  \label{fig:fig1tv}
  \end{figure*}

\subsection{Constant uncertain parameter}

Denote by $\th$ an uncertain real parameter in the drift term of the
single qubit Hamiltonian system,
\beq[eq:ex H const]
H_t = v_t\sig_x + (1+\th)\sig_z
\quad
\left\{
\bea{l}
\Hb_t = v_t\sig_x + \sig_z
\\
\Ht_t = \th\sig_z,\
-\del\leq\th\leq\del
\eea
\right.
\eeq
where $v_t,\tN$ are the $N$ piece-wise-constant (PWC) control pulse
magnitudes of the (single) control variable over the normalized time
interval $[0,1]$ with uniform pulse widths $1/N$. Noting that the
parameter uncertainty belongs to the class $\vec\th\in\Theta_\infty$
from \refeq{Jparset}, the form of the objective function follows from
\refeq{Jgen} as,
\beq[eq:J ex1]
\Obj = \left\{1-\Fnom,\ \normsm{\Acal}_\infty\right\}\
\left\{
\bea{l}
\Fnom = |\trace(W^\dag\Ub_N/2)|^2
\\
\Acal=\mat{\vec A_1&\cdots&\vec A_M}
\\
A_t = \Ub_t^\dag\sig_z\Ub_t/M
\eea
\right.
\eeq
Since there is a single parameter, the bound $\del$ is not needed in
the robustness measure, only for evaluation.

\figref{fig1const}(a)-(b)-(c) shows the numerical results using the
optimization procedure in \refeq{Fup}-\refeq{Jdn}. The goal is to make
a robust identity gate ($W=I_2$) with $N=5$ PWC control pulses and
with averaging over $M=50$ uniform samples.  For evaluation the
uncertain constant parameter range is bounded by $\del=0.05$. In the
first stage of the algorithm \refeq{Fup}, the fidelity threshold is
set at $f_0=1-10^{-6}$. During the second stage \refeq{Jdn} the
weighting parameter which trades control increment size against
robustness decrease (\refeq{singquad0} is set at
$\alf=5$.

\figref{fig1const}(a) shows the expected behavior: in the first stage
only fidelity error (blue) is minimized. At 37 iterations the
threshold $f_0=1-10^{-6}$ is reached and robustness minimization (red)
begins. In the second stage while the fidelity error fluctuates (blue)
around $1-f_0$ (these only look large on the log scale) the robustness
measure decreases eventually converging to a small
value. \figref{fig1const}(b) compares the fidelity error vs. uncertain
parameter level with the stage-1 optimal control (blue plot at 37
iterations) against the robust control (red at final iteration); the
latter exhibiting almost two-orders of magnitude improvement at the
extremes of the parameter variation: about 0.99999 in fidelity at
$\th=\pm 0.05$.

\figref{fig1const}(a) -- and in other examples to follow -- the
topological property of the quantum control landscape at or near the
top of the landscape is revealed. Specifically, at high values of
fidelity, though the fidelity gradient is very small, the Hessian
easily allows ``roaming at the top of the landscape.'' The result is
that the nominal fidelity stays high while the robustness measure is
decreased. This is equivalent to manipulating the unitary trajectory
so that fidelity remains high, at or near the top of the landscape, so
that the additional control freedom in the available null space can
provide robustness \toprefs.

A visual inspection of the two pulse sequences in
\figref{fig1const}(c) offers little insight into why the second
sequence produces a significant level of robustness. Though the
magnitudes of the robust sequence is much larger, it is not understood
at this time what is the \emph{precise} mechanism which achieves such
a significant robustness advantage. The basis for understanding
provided by Theorem~\ref{thm:avg} is somewhat coarse: make the
time-averaged Hamiltonian small, which clearly occurs.

\subsection{Additive Hamiltonian parameter noise}

For this example we replace the \emph{constant} uncertain parameter
$\th$ in \refeq{ex H const} by $\th_t,\tM$, a real \emph{time-varying}
parameter resulting in the Hamiltonian,
\beq[eq:ex addnoise]
H_t = v_t\sig_x + (1+\th_t)\sig_z
\eeq
We again set $N=5$ PWC control pulses, $M=50$ noise samples, and the
identity target. Here we take the noise to be uniformly random in $\pm
0.05$ with \emph{two} magnitude changes over the time interval. This
noise has a ``sinc'' spectrum \refeq{Fom ell}. The resulting
robustness measure is given by \refeq{J thell det}.

\figref{fig1tv}(a)-(c) displays the simulation results. The fidelity
error and robustness measure plots in \figref{fig1tv}(a) follow the
previous pattern of \figref{fig1const}(a); here the switching from fidelity
error minimization to robustness minimization occurs at iteration 27
when fidelity exceeds $f_0=1-10^{-5}$. \figref{fig1tv}(b) compares the
fidelity range with the control at switching with the final robust
control; the plots show the mean and deviations for 100 random samples
at each selected uncertainty magnitude in the range out to $\pm
0.05$. \figref{fig1tv}(c) shows the control pulses at switching and
final.
Clearly a considerable level of robustness is achieved, over an order
of magnitude. Interestingly the robust pulse sequence exhibits a
symmetric pattern which suggests some possibly generic means to
overcome this type of time-varying uncertainty.

\section{Concluding Remarks \& Outlook}
\label{sec:sum}

The approach to robust quantum control presented here rests on two
theoretical foundations: the classic method of averaging
\cite{Hale:80} and the known topological properties of the quantum
control landscape \toprefs.  Application of the method of averaging
directly results in a multicriterion optimization problem consisting
of the uncertainty-free fidelity competing with a generic robustness
measure, the latter being a time-domain product of a function (or
functional) of the controls and the character of the uncertainty
\refeq{Jgen}.  Such a product form is ubiquitous in the many
applications of the \emph{Small Gain Theorem}
\cite{Zames:1966,DesVid:1975} and generalizations for various
uncertainty models, \eg, \cite{LMI94,ZhouDG:96}. These require a
product in a feedback loop to be small: an uncertainty bound and an
uncertainty-free (closed-loop) map.  From the averaging theory there
naturally arises a quantitative norm measure of a time-averaged
Hamiltonian \refeq{Gavg} reflective of the interaction between the
uncertain perturbation and the uncertainty-free unitary
evolution. This is the first term in a Magnus expansion of the
interaction representation between the nominal unitary and the
uncertainty. If this term is sufficiently small, then at most, only
second order error effects can accrue \refeq{rbst}.  In the case of
quantum information sciences for established realizations, the tacit
assumption is that the (possibly known or unknown) perturbations are
small.


The second theoretical foundation, properties of the control landscape
topology, motivated a \emph{two-stage algorithm} \refeq{twostage} for
finding the robust control: first, maximize fidelity without regard to
uncertainty, then, at a high fidelity threshold, switch to
maximizing robustness while maintaining the high uncertainty-free
fidelity level. \black{We view the time-domain robustness
  measures developed here together with the two-stage algorithm to
  solve the multicrterion problem to be complementary to the
  frequency-domain filter function measures and approaches in \ffrefs.}

In Appendix \S\refsec{comp} we showed that by using a
quadratic approximation of \emph{both} fidelity and robustness in this
region of the control landscape, the task of finding a control
increment at each iteration is a convex optimization problem. The
actual system fidelity robustness of course is subject to the accuracy
of the quadratic approximations.

We derived the specific form of the matrix products in the robustness
measure corresponding to many common uncertainties.  The examples
presented, using our two-stage algorithm, clearly illustrate the known
flexibility inherent in the landscape topology at high fidelity.
However, a visual inspection of many of the robust control pulse
sequences does not offer direct insight into why a significant level
of robustness is achieved, \ie, the mechanism involved. In a few cases
there are discernible symmetric pulse patterns. Unfortunately the
basis for understanding provided by Theorem~\ref{thm:avg} is somewhat
coarse: make the time-averaged Hamiltonian small, \eg, via
\refeq{Jgen}, and yet was shown to operate well. However, there is no
guarantee that any particular degree of robustness can be achieved,
and the best control action at the top of the landscape would most
likely call for a global search, unlike the local procedure utilized
in this paper.

It is important to emphasize that the \emph{actual} effectiveness of
the resulting optimization as predicted by the averaging theory, or by
any other approach, can only be determined by evaluation on the total
system with uncertainties included. For either deterministic or
probabilistic uncertainty sets, this requires simulating the full
system with a sufficient number of samples from the uncertainty
sets. Ultimately an experiment is the final proof.

Looking ahead, we see a significant unanswered question: \emph{How is
  robustness precisely achieved for a quantum system?} What is the
underlying mechanism? We mentioned the close connection between the
generic form of the robustness measure \refeq{Jgen} and the Small Gain
Theorem \cite{Zames:1966,DesVid:1975}. For robust linear control the
latter (and subsequent variations \cite{LMI94,ZhouDG:96}) have proven
to be extraordinarily effective (and capable of revealing the
robustness mechanism) even against a large class of
uncertainties. Moreover, because the mechanism is understood, and in
many cases, comprehensible, it is possible to describe a very large
class of uncertainty models which encompass what is anticipated
without having to delineate each specific possibility
\cite{id4c:92}. Why not for quantum?
%
For example, is it possible to build a ``standard'' set of uncertainty
models, each of which captures a large variety of possibilities,
thereby leading to a universal set of robust quantum controls?  A
qualitative generic answer can be understood in terms of the landscape
``roaming'' feature, \ie, seeking a control where the disturbance has
minimal eigenvector projection along the Hessian eigenvectors of
negative value. Is this what allows a quantitative understanding of
the favorable solutions of the dual problem \refeq{dual0},
\refeq{dual} such that maximum robustness is achieved at high
fidelity?  Notwithstanding the latter research direction, the
fundamental quantum robustness challenge stated above remains.

\section*{Acknowledgments} 

RLK, GB, and HR acknowledge support from the U.S. Department of Energy
(DOE) under STTR Contract No.  DE-SC0020618.  Additionally, HR
acknowledges support from the ARO under Contract
No. W911NF-19-0382. The authors thank Daniel Lidar, Jason Dominy, and
Robert Bitmead for helpful discussions.
%
%
\bibliographystyle{unsrt}
\bibliography{rlk}

\begin{appendix}
\flushbottom
\section{Proof of Theorem \ref{thm:avg}}
\label{sec:avg analysis}

With no loss in generality, set the bounded Hamiltonian perturbation
in \refeq{Hcalunc} to $\Ht(t)=\del\B(t)$ where for a specified matrix
norm, $\normsm{\B(t)}=1$.  The interaction unitary $R(t),t\in[0,T]$ from
\refeq{Rt} then evolves from,
\beq[eq:R]
\beasep{1.25}{rcl}
i\dot R(t) &=& \del A(t)R(t),\ R(0)=I_n
\\
A(t) &=& \Ub(t)^\dag\B(t)\Ub(t)
\eea
\eeq
where $\Ub(t)$ is the nominal (error-free) unitary
  \refeq{Unom}. Define the average interaction unitary,
\beq[eq:Rbar]
\Rb(T) = e^{-i\del T\avg{A}},
\quad
\avg{A} = \frac{1}{T}\int_0^TA(t)dt
\eeq
%
To prove Theorem
  \ref{thm:avg} first requires showing that the interaction unitaries
  satisfy the following inequalities:
\beq[eq:interaction]
\beasep{1.5}{l}
\norm{\bar R(T)-I} \leq e^{\gamb(\del T)}-1
\\
\norm{R(T)-\bar R(T)} \leq e^{\gamt(\del T)^2}-1
\eea
\eeq
with $(\gamb,\gamt)$ as defined in \refeq{rbst} of Theorem
\ref{thm:avg}, that is,
\beq[eq:gams]
\gamb = \normsm{\avg{A}},
\quad
\gamt = \max_t\normsm{A(t)-\avg{A}}
\eeq
Using the form of the standard state transformation for averaging
analysis described in \cite[\S V.3]{Hale:80} (periodicity, usually
assumed, is not needed here) set,
\beq[eq:r2v]
\beasep{1.25}{rcl}
R(t) &=& (I+\del K(t))V(t)
\\
K(t) &=& -i\int_0^t\left(A(\tau)-\avg{A}\right)d\tau
\eea
\eeq
The matrix $V(t)$, which is is not necessarily a
unitary, is the solution of,
\beq[eq:v]
\bea{rcl}
i\dot V &=& \del\left( \avg{A}+\D(t) \right)V,\
V(0)=I_n
\\
\D(t) &=& \del(I+\del K(t))^{-1}(A(t)K(t)-K(t)\avg{A})
\eea
\eeq
With $\normsm{A(t)}\leq 1$, $\normsm{\avg{A}}\leq 1$, and
$\normsm{A(t)-\avg{A}}\leq\gamt$, it follows that
$\norm{K(t)}\leq\gamt t$. Additionally, $K(t)+K(t)^\dag=0$ ensures
that $\normsm{(I+\del K(t))^{-1}}\leq 1$. As a result,
\beq[eq:Dbnd]
\beasep{1.25}{rcl}
\normsm{D(t)} &\leq& \del\normsm{A(t)K(t)-K(t)\avg{A}}
\\
&\leq& 2\del\normsm{K(t)}
\leq 2\del\gamt t
\eea
\eeq
Applying \emph{variation of constants} to \refeq{v} gives,
\beq[eq:var1]
\beasep{1.25}{rcl}
V(t) &=& \Rb(t) -i\del\Rb(t) 
\int_0^t \Rb(\tau)^\dag\D(\tau)V(\tau)d\tau
\eea
\eeq
Introducing the error,
\beq[eq:err]
E(t) = V(t)-\Rb(t)
\eeq
Substituting into \refeq{var1} gives, 
\beq[eq:var2]
E(t)=
-i\del\Rb(t)\int_0^t
\Rb(\tau)^\dag\D(\tau)\left(\Rb(\tau)+E(\tau)\right)d\tau
\eeq
Since $\Rb(t)$ is a unitary, $\normsm{\Rb(t)}=1$. This together with
the previous bound on $D(t)$ gives,
\beq[eq:ebndt]
\beasep{1.5}{rcl}
\norm{E(t)} &\leq&
2\gamt\del^2\left(\int_0^t\tau d\tau + \int_0^t\tau\norm{E(\tau)}d\tau\right)
\\
&=&
\gamt(\del t)^2 + 2\gamt\del^2\int_0^t\tau\norm{E(\tau)}d\tau
\eea
\eeq
Applying the Bellman-Gronwall Lemma together with the fact that
$K(T)=0$ implies $R(T)=V(T)$, and hence, $E(T)=R(T)-\bar R(T)$, yields
the second inequality in \refeq{interaction}. The first inequality in
\refeq{interaction} is a known result which can also be obtained
by following the same argument which
resulted in the first inequality, now substituting the identity for
$R(t)$ and then applying Bellman-Gronwell.

To prove the main result \refeq{rbst}, we need the following
relationship established in \cite{KosutArenzRabitz:2019} between
fidelity and the Frobenius norm:
\beq[eq:fidnorm]
\beasep{1.5}{l}
\Ecal=\min_{|\phi|=1}
\normsmf{U(T)-\phi W}^2 = 2n\left(1-\sqrt{F}\right)
\\
F = |\trace(W^\dag U(T)/n)|^2
\\
\phi = \trace(W^\dag U(T)/n)/|\trace(W^\dag U(T)/n)|
\eea
\eeq
Likewise for the nominal fidelity,
\beq[eq:fidnom norm]
\beasep{1.5}{l}
\Ecal_\nom = \min_{|\phi|=1}
\normsmf{\Ub(T)-\phi W}^2 = 2n\left(1-\sqrt{\Fnom}\right)
\\
\Fnom = |\trace(W^\dag \Ub(T)/n)|^2
\\
\widebar\phi = \trace(W^\dag \Ub(T)/n)/|\trace(W^\dag \Ub(T)/n)|
\eea
\eeq
To simplify notation we drop the final time $T$ in all the variables,
\eg, let $U=U(T),\Ub=\Ub(T)$, \etc. Applying
\refeq{fidnorm}-\refeq{fidnom norm} yields,
\beq
\beasep{1.5}{rl}
\Ecal &\leq \normsmf{U-\bar\phi W}^2
\\
&= \normsmf{\Ub-\bar\phi W + \Ub(R-I)}^2\
\mbox{( using $U=\Ub R$ )}
\\
&\leq
\left(\sqrt{\Ecal_\nom}+\normsmf{R-I}\right)^2\
\mbox{( from \refeq{fidnom norm} )}
\eea
\eeq
From the definitions of $\Ecal$ and $\Ecal_\nom$, the last inequality
above is equivalent to,
\beq[eq:Fbnd]
\beasep{1.5}{rl}
F &\geq \Fnom - (\widetilde\Ecal/n)\sqrt{\Fnom}
+\left(\widetilde\Ecal/2n\right)^2
\\
\widetilde\Ecal &=
2\sqrt{\Ecal_\nom}\normsmf{R-I}+\normsmf{R-I}^2
\eea
\eeq
If $\Fnom\approx 1$, then $\Ecal_\nom\approx 0$,
$\widetilde\Ecal\approx\normsmf{R-I}^2$ and,
\beq
F \geq 1 - \normsmf{R-I}^2/n + \left(\normsmf{R-I}^2/2n\right)^2 
\eeq
Thus,
\beq
\beasep{1.5}{rl}
F
&= 1-\Ocal\{\norm{R-I}^2\}
\\
&= 1-\Ocal\{\norm{\widebar R-I + R-\widebar R}^2\}
\\
&=
1-\left(\Ocal\{\norm{\widebar R-I}\}
+ \Ocal\{\norm{R-\widebar R}\}\right)^2
\\
&= 1 - \left(\Ocal\{\gamb(\del T)\} + \Ocal\{\gamt(\del T)^2\}\right)^2\
\mbox{(from \refeq{interaction})}
\eea
\eeq
The last line establishes Theorem \ref{thm:avg}. 

\section{Two-stage algorithm}
\label{sec:comp}
With $m$ PWC controls over $N$ uniform intervals, the two-stage
algorithm in \S~\refsec{twoalg} to find the control variable
magnitudes $v\in\Rbf^\mc$ can be summarized as follows:

\ben\setlength{\itemsep}{0.05in}
\item {\bf initialize} $v\in\Vcalcon$, set a fidelity threshold
  $f_0\approx 1$ and a control increment bound $\bet$.
\item {\bf repeat} for $i=1,2,\ldots$ until satisfied
  \ben\setlength{\itemsep}{0.05in}
\item {\bf while} $\Fnom(v^i) < f_0$, {\bf do}
  \beq[eq:Fup]
  v^{i+1} = \arg\min_v
  \left\{1-\Fnom(v)\ \Bigg|\
  \bea{l} v = v^i + \vinc \in \Vcalcon
  \\
  \norm{\widetilde v}\leq \bet
  \eea
  \right\}
  \eeq

\item {\bf while} $\Fnom(v^i) \geq f_0$, {\bf do}
  \beq[eq:Jdn]
  v^{i+1} = \arg\min_v
  \left\{\Jrbst(v)\ \Bigg|\
  \bea{l} v = v^i + \vinc \in \Vcalcon
  \\
  \norm{\widetilde v}\leq \bet
  \\
  \Fnom(v) \geq f_0
  \eea
  \right\}
  \eeq

  \item {\bf stop} when $\Jrbst(v^{i+1})\approx\Jrbst(v^i)$
\een
\een
Since the objective switches from minimizing the uncertainty-free
fidelity error, $1-\Fnom(v)$ to minimizing the robustness measure
$\Jrbst(v)$ at the fidelity threshold $f_0\approx 1$, it is expected
that the accompanying fidelity gradient $\nabla\Fnom(v)$ will be
small. Consequently, the fidelity is best approximated as a quadratic
function of the control increment $\vinc$ in the neighborhood of the
current control $v$. That is, for a sufficiently small $\bet$ with
$\norm{\widetilde v}\leq\bet$ the nominal fidelity is approximated as,
\beq[eq:Fquad]
\bea{rcl}
\Fnom(v+\widetilde v) &\approx& \Fquad(\vinc)
\\
&=& \Fnom+
\nabla\Fnom^T\vinc
+\vinc^T\nabla^2\Fnom\vinc/2
\eea
\eeq
where $\Fnom,\nabla\Fnom,\nabla^2\Fnom$ are evaluated at $v$, the
current control. The Hessian $\nabla^2\Fnom$ may not be negative
definite (or negative semidefinite) as would be the case at a global
fidelity maximum $v^\star$ where $\Fnom(v^\star)=1$ and
$\nabla\Fnom(v^\star)=0$. However, it is known from topological
arguments that at the top, at any global optimum $v^\star$, the
Hessian $\nabla\Fnom(v^\star)$ has $(\mc)^\star<\mc$ negative
eigenvalues, the rest are zero, and hence, it is negative
semidefinite.

With $\Fnom\approx 1$, it is to be expected that some number of
eigenvalues of $\nabla^2\Fnom$ are near zero. This motivates replacing
$\nabla^2\Fnom$ with an approximation obtained by setting the small
eigenvalues to \emph{exactly} zero. Any control in the null-space
defined by the corresponding $\mc-(\mc)^\star$ eigenvectors can be
used for purposes other than keeping $\Fnom(v)\geq f_0$; in
particular, minimizing the robustness measure.  However, the solution
method we propose here does not require the Hessian to be positive
semidefinite, so there is no need to make a further approximation of
the fidelity Hessian.

In the neighborhood of the current control $v$, we also approximate
the robustness measure as a quadratic in the control increment
$\vinc$, as follows:
\beq[eq:Jquad]
\bea{rcl}
\Jrbst(v+\widetilde v) &\approx& \Jquad(\vinc)
\\
&=& \Jrbst +
\nabla\Jrbst^T\vinc+\vinc^T\nabla^2\Jrbst\vinc/2
\eea
\eeq
where $\Jrbst,\nabla\Jrbst,\nabla^2\Jrbst$ are evaluated at $v$, the
current control.  Even though the fidelity gradient $\nabla\Fnom$ is
small when $\Fnom\approx 1$, there is no reason to expect that the
robustness gradient, $\nabla\Jrbst$, is small, nor that its Hessian,
$\nabla^2\Jrbst$, is positive semidefinite. Again, we make no further
assumptions or approximations.

In Stage 2, having approximated both fidelity and robustness
objectives as a quadratic in control increment $\vinc\in\Rbf^\mc$ via
\refeq{Fquad}-\refeq{Jquad}, we seek the largest decrease in
robustness, that is, the largest $\Del>0$ such that there exists
$\vinc\in\Rbf^\mc$ simultaneously satisfying the three constraints:
  \beq
  \vinc^T\vinc\leq\bet,\quad
  \Fquad(\vinc)\geq f_0,\quad
  \Jquad(\vinc) \leq \Jrbst - \Del
  \eeq
Not all these parameters are known: in general would like $\bet$ to be
small to satisfy the quadratic approximations and $\Del$ as large as
possible. The threshold fidelity can be set to an $f_0$ near 1, how
 ``near'' to 1 may vary per problem. The trade-off amongst the
three goals can be expressed by the constrained optimization,
  \beq[eq:singquad0]
  \beasep{1.5}{ll}
  \mbox{minimize}& \vinc^T\vinc/2 + \alf(\Jquad(\vinc)-\Jrbst)
  \\
  \mbox{subject to}&
    \Fquad(\vinc)\geq f_0\quad
  \eea
  \eeq
The objective simultaneously reflects the first two goals where
$\alf>0$ weighs the robustness decrease $(-\Del)$ relative to the
control increment size.  Thus the values for $\bet$ and $\Del$ are
implicit in the objective; only $f_0$ and $\alf$ are explicitly
set. Using the quadratic approximations, the optimization
\refeq{singquad0} becomes,
    \beq[eq:singquad1]
    \beasep{1.5}{llll}
    \mbox{minimize} &
    \vinc^T(I+\alf\nabla^2\Jrbst)\vinc + 2\alf\nabla\Jrbst^T\vinc
    &
    \\
    \mbox{subject to} &
    \vinc^T\nabla^2\Fnom\vinc +2\nabla\Fnom^T\vinc +2(\Fnom-f_0) \geq 0
    &
    \eea
    \eeq
At the current control $v$,
\refeq{singquad1} is not guaranteed to be a convex optimization
problem in the increment $\vinc$.  However, as shown in
\cite{BoydV:04}, the dual problem in the two variables
$\lam,\gam\in\Rbf$ can be obtained from the linear-matrix-inequality
(LMI) constrained convex optimization,
      \beq[eq:dual]
      \beasep{1.25}{l}
      \mbox{maximize}\quad \gam
      \\
      \mbox{subject to}\quad
      \lam\geq 0
      \\
      \mat{I-\lam\nabla^2\Fnom+\alf\nabla^2\Jrbst
        &\alf\nabla\Jrbst-\lam\nabla\Fnom
      \\
      (\alf\nabla\Jrbst-\lam\nabla\Fnom)^T&-2\lam(\Fnom-f_0)-\gam} \geq 0
    \eea
    \eeq
Having obtained $\lam$, the resulting optimal solution for the control
increment is,
    \beq[eq:vinc opt]
    \vinc = -\left(I-\lam\nabla^2\Fnom
    +\alf\nabla^2\Jrbst\right)^{-1}(\alf\nabla\Jrbst-\lam\nabla\Fnom)
    \eeq
From \cite{BoydV:04}, if there exists an increment $\vinc$ such that
the constraint is a strict inequality, that is, $\Fquad(\dv)>f_0$,
then \emph{strong duality} holds and the two objectives in
\refeq{singquad1} and \refeq{dual} are equal, \ie,
    \beq
    \vinc^T(I+\alf\nabla^2\Jrbst)\vinc+2\alf\nabla\Jrbst^T\vinc=\gam
    \eeq
Since $\Fnom>f_0$ and if the LMI in \refeq{dual} is positive definite, it
follows that $\gam < -2\lam(\Fnom-f_0)$. Thus $\gam$ is (theoretically)
always negative which ensures that the quadratic approximation
$\Jquad(\vinc)$ of $\Jrbst(v+\vinc)$ decreases. As we see in the
numerical examples, the size of the decrease in $\Jrbst(v+\vinc)$
depends on the choice of the weighting parameter $\alf$ in the
objective \refeq{singquad0}. In addition, since we are roaming about
near the top of the landscape where fidelity is very near to 1, the
constraint that $\Fquad(\vinc)\geq f_0$ is easily violated by very
small amounts. These look large on the log-scale plots of fidelity
error vs. iteration.

\section{Uncertainty Models}
\label{sec:more uncmod}


\subsection{Linear time invariant noise}

Although \refeq{BLT} captures \emph{all} LTV filters, the overall
character of the noise will dictate some specific forms. For example,
suppose that it is only known that the noisy parameter has a limited
bandwidth. The least complicated system which captures the bandwidth
constraint only, is a first-order linear-time-invariant (LTI) system
with time constant $\bet$ (bandwidth is effectively $1/\bet$) and
whose input, sampled at $T/M$, is the stationary zero-mean white noise
sequence $\vec w,\cov(\vec w)=\del_w^2I_M$. Then $\th_t,\tM$ is the
output of the discrete-time \emph{zero-order-hold} system with
continuous-time transfer function $1/(\bet s+1)$, \ie,
$\Theta(z)=(1-a)/(z-a)W(z)$ with $a=\exp\{-\frac{T/M}{\bet}\}$.
It follows that the $M\times M$ filter matrix $\Kcal$ in \refeq{BLT} is
\beq[eq:S lti]
\Kcal = (1-a)\mat{1&&&\\a&1&&\\
    \vdots&\ddots&\ddots&\\
    a^{M-1}&\cdots&a&1}
\eeq
The discrete-time Fourier transform $\Theta(\om)$ of $\th_t,\tM$ is
bounded by,
\beq[eq:Fom]
|\Theta(\om)| \leq \del
\frac{1-a}{\sqrt{2(1-\cos(\om T/M))+a^2}}
\eeq
%
  
\subsection{PWC noise}

A more extreme type of uncertain time-variation is where $\th_t$ is
piece-wise-constant. Suppose $\th_t$ is constant but bounded
over $L$ uniform intervals $\Tcal_\ell$, each of width $T/L$ in
$[0,T]$. Thus $\th_t$ changes abruptly $L$ times (jumps) from a given
underlying distribution. Suppose for each of the $L$ intervals the
uncertain parameter $\th_t,t\in\Tcal_\ell$ is independently bounded by
$\norm{\th_t,t\in\Tcal_\ell}_\infty\leq\del_\infty$. In other words,
consider the \emph{deterministic noise uncertainty set},
\beq[eq:Hcal thellt]
\Hcalunc=\left\{
\Ht_t=\th_t\B_t\ 
\left|
\bea{l}
\th_t = w_\ell,\ t\in\Tcal_\ell,\ \ell=1:L
\\
\norm{\th_\ell}_\infty\leq\del_\ell
\eea
\right.
\right\}
\eeq
This type of uncertainty has an equivalent frequency domain
spectrum. Since in each of the $L$ time intervals the uncertain
parameter varies independently, it can be considered as a
time-sequence over $[0,T]$ of uncertain amplitude $\th_\ell$ with a
fixed duration $T/L$, \ie, a pulse of known duration and uncertain
magnitude bounded by $\del$. For a deterministic uncertainty
\refeq{Hcal thellt}, the magnitude of the Fourier transform
$\Theta(\om)$ of such a sequence is bounded by the ``sinc'' function,
\ie, for each $\ell=1:L$,
\beq[eq:Fom ell]
|\Theta_\ell(\om)| \leq \del 
\left(\frac{2\pi}{\om_L}\right)
\left|
\sinc\left(\frac{\om\pi}{\om_L}\right)
\right|,\
\om_L = \frac{2\pi}{T/L}
\eeq
As the number of intervals increase the spectral magnitude decreases
while the bandwidth ($\sim \om_L$) increases, that is, the frequency
spreads which potentially makes robustness more difficult to
attain. The needed temporal behavior of $\Ub_t$ to overcome the
frequency spread may be harder to realize under control limitations.
The corresponding vector of the time-averaged Hamiltonian is,
\beq[eq:Havg thell]
\vecGavg = \Acal S\vec w\quad
\left\{
\beasep{1.5}{l}
\Acal = \mat{\vec \Acal_1&\cdots&\vec \Acal_M} 
\\
S = {\rm blk\_diag}\mat{\un_{M/L}&\cdots&\un_{M/L}}
\\
\Acal_\ell = (1/M)\sum_{t\in\Tcal_\ell}\Ub_t^\dag B_t \Ub_t
\eea
\right.
\eeq
where $\un_{M/L}$ is an $M/L$ vector of ones, $\Acal$ is $n^2\times
M$, $S$ is $M\times L$, and $\norm{\vec w}_\infty\leq\del$.  As a
consequence, the robustness measure is the induced matrix norm,
\beq[eq:J thell det]
\Jrbst = \del\norm{\Acal S}_\infty
\eeq
In effect, each of the piece-wise-constant uncertain parameters is
reflected by a corresponding time-averaged Hamiltonian over each
interval $\Tcal_\ell$ resulting in a collection of uncertain constant
bounded parameters each of which is held fixed over some time
intervals. For a probabilistic uncertainty where in each of the $L$
intervals $\th_t$ is zero-mean with covariance $\del^2 I_L$, the
corresponding robustness measure becomes,
\beq[eq:J thell prob]
\Jrbst = \del\normf{\Acal S} 
\eeq
%


\subsection{Robustness measure as filter function}
\label{sec:fltr func}

\black{The robustness measure derived from the averaging approach is
equivalent in at least one instance to a \emph{filter function}
measure as described in \ffrefs.}
%
%
As an illustration, assume that the interaction Hamiltonian \refeq{Rt}
is $G(t)=\th(t)A(t)$ with $A(t) = \Ub(t)^\dag\B(t)\Ub(t)$, where
$B(t)$ is known and where $\th(t)$ is a zero-mean wide-sense
stationary random process with autocorrelation function,
$s(\tau) = \Ebf\{\th(t)\th(t+\tau)\}$.
Clearly $G(t)$ and the time-averaged Hamiltonian
$\avg{G}=(1/T)\int_0^TG(t)dt$ are both zero mean. Using the Frobenius
norm to reflect the size of $\avg{G}$, let the robustness measure be
the variance, \ie, $\Jrbst = \Ebf\normsmf{T\avg{G}}^2$ where expectation
is with respect to the noise $\th(t)$. The result is,
\beq[eq:Jff0]
\beasep{1.5}{rl}
\Jrbst
&= \int_{t=0}^T\int_{t'=0}^T
s(t-t')c(t,t')
dt' dt
\\
c(t,t') &= \trace\left(\Ub(t-t')^\dag\B(t)\Ub(t-t')\B(t')\right)
\eea
\eeq
Replace $s(t-t')$ above by using the inverse Fourier transform
$s(\tau) = (1/2\pi)\int_{-\infty}^\infty S(\om)e^{i\om\tau}d\om$ where
$S(\om)$ is the associated power spectral density of the noise. This
gives,
%
\beq[eq:Jff1]
\Jrbst
= \int_{-\infty}^\infty {\cal A}(\om)S(\om)d\om
\eeq
%
with,
\beq[eq:Jff2]
{\cal A}(\om) = \frac{1}{2\pi}
\int_{t=0}^T\int_{t'=0}^T e^{i\om(t-t')}c(t,t')dt'dt
\eeq
If in \refeq{Jff0}, the known part of the uncertain perturbation is
$B(t)=B$, a constant, then $c(t,t')=c(t-t')=c(t'-t)$ which gives,
\beq[eq:Jff3]
\beasep{1.5}{rl}
{\cal A}(\om) &= \frac{1}{\pi}
\int_{0}^T (T-\tau)e^{i\om\tau}c(\tau)d\tau
\\
c(\tau) &= \trace\left(\Ub(\tau)^\dag\B\Ub(\tau)\B\right)
\eea
\eeq
Sometimes $S(\om)$ can be measured, and even then, it can be
approximated by \refeq{Fom} and/or \refeq{Fom ell}. The frequency
domain actuator dynamics error \refeq{Dom} is also a candidate.

\subsection{Additive control noise}
\label{sec:add noise}

Suppose that each of the $m$ controls $v_{jt},j=1:m$ are affected by
\emph{additive noise} $\{\th_{jt},\tM\}$. As a result,
$v_{jt}=(\vb_{jt}+\th_{jt})H_c^j$ where $\vb_{jt},j=1:m$, the
disturbance-free controls, serve as the design variables. As an
illustration, suppose that the additive noise in control $j$, denoted
by the $M$-vector $\vec\th_j$, is the output of a filter $\Kcal_j$ as
in \refeq{LTV}-\refeq{BLT} with $\vec w_j$ being a zero-mean sequence
with $\cov(\vec w_j)=\del_j^2I_M$. This is the same ``model'' of
uncertainty as given by \refeq{J thtv mult} except here we replace the
time-varying $B_{\ell t}$ with the constant control Hamiltonians
$H_c^j$. The additive control noise Hamiltonian uncertainty set is,
\beq[eq:Hcal thadd]
\Hcalunc=\left\{
\Ht_t=\sum_{j=1}^m\th_{jt}H_c^j,\ 
\left|
\bea{l}
\vec\th_j=\Kcal\vec w_j,\ \mbox{$\vec w_j$ zero-mean}
\\
\cov(\vec w_j)=\del_w^2I_{M}
\eea
\right.
\right\}
\eeq
Following the same procedure to arrive at \refeq{J thtv mult}, the
robustness measure for this model of additive control noise is,
\beq[eq:J thadd]
\Jrbst = \sum_{j=1}^m
\del_j\normf{\Acal_j  S_j}
  \quad
  \left\{
  \bea{l}
  \Acal_j=\mat{\vec A_{j 1}&\cdots&\vec A_{j M}}
  \\
  A_{j t} = \Ub_t^\dag B_{j t}\Ub_t
  \\
  S_j = \Kcal_j/\normf{\Kcal_j}
  \eea
  \right.
  \eeq
where each $\Acal_j$ is $n^2\times M$.  Here we assumed that the noise
affecting each control is independently drawn from different
independent distributions each with its own spectrum (via
$\vec\th_j=\Kcal_j\vec w_j$). Though there are too many variations to
delineate, such as noise drawn independently with the same spectrum,
\eg, $\vec\th_j=\Kcal\vec w_j$, or combinations thereof, the procedure
to arrive at the appropriate robustness measure is the same. Moreover,
the form of the robustness measure in all these cases will conform to
the generic representation given by \refeq{Jgen}.

\subsection{Multiplicative control noise}

Suppose each of the $m$ controls $v_{jt},j=1:m$ is affected by a
time-varying \emph{multiplicative noise} function $\th_{jt}$ such that
$v_{jt}=(1+\th_{jt})\vb_{jt}H_c^j$ where, as previously described,
$\vb_{jt},j=1:m$ are the disturbance-free controls. As in the additive
noise example, assume that for each control the noise is drawn
independently from the same distribution. The associated Hamiltonian
uncertainty set is almost identical to \refeq{Hcal thadd} except for
the dependence on the control:
\beq[eq:Hcal thmult]
\Hcalunc=\left\{
\Ht_t=\sum_{j=1}^m\th_{jt}\vb_{jt}H_c^j,\ 
\left|
\bea{l}
\vec\th=\Kcal\vec w,\
\mbox{$\vec w$ zero-mean}
\\
\cov(\vec w)=\del_w^2I_{mM}
\eea
\right.
\right\}
\eeq
Following the procedure to obtain \refeq{J thadd},
%
\beq[eq:J thmult]
\Jrbst = \sum_{j=1}^m
\del_j\normf{\Acal_j  S_j}
  \quad
  \left\{
  \bea{l}
  \Acal_j=\mat{\vec A_{j 1}&\cdots&\vec A_{j M}}
  \\
  A_{j t} = \vb_{jt}\Ub_t^\dag B_{j t}\Ub_t
  \\
  S_j = \Kcal_j/\normf{\Kcal_j}
  \eea
  \right.
  \eeq
The only difference between this measure and \refeq{J thadd} is that
each disturbance-free control weighs the control Hamiltonians.

\subsection{Uncertain actuator dynamics}
\label{sec:unc act}

Though not strictly noise, uncertain actuator (control generator)
dynamics can appear in the form of a multiplicative control
perturbation.  Suppose the actuator dynamics are well modeled as an
LTI system such that each of the $m$ control commands $\vb_{tj}$ are
effected by similar dynamics, \ie,
\beq[eq:vb2v]
v_{tj} = D_j(z)\vb_{tj},\ j=1:m,\ t=1:M
\eeq
where $D_j(z)$ is a rational transfer function in the unit delay
operator $z^{-1}$ corresponding to the sampling rate $T/M$. Assume
that $D_j(z)$ is in the uncertainty set,
\beq[eq:unc act]
{\cal D}(\Del) = \left\{
D(z)=(1+\Del(z)Q(z))\bar D(z)\
\left|\
\bea{l}
\hinfnorm{\Del}\leq\del
\\
\hinfnorm{Q}=1
\eea
\right.
\right\}
\eeq
The transfer function $\bar D(z)$ is the perturbation-free LTI model
of the actuator dynamics, $Q(z)$ is a normalized weighting factor, and
$\Del(z)$ is a completely uncertain transfer function bounded in
$\hinf$ norm by $\del$, the maximum magnitude of the frequency
response. This set characterizes a standard transfer function with a
bounded multiplicative model-error \cite{ZhouDG:96}. The equivalent
frequency domain constraint is,
\beq[eq:Dom]
|D(i\om)/\bar D(i\om)-1|\leq\del|Q(i\om)|
\eeq
Since $Q(z)$ is normalized, the left hand side can never exceed $\del$
over all frequencies.

The nominal and error Hamiltonians associated with this class of
uncertainty are,
\beq[eq:ham unc act]
\beasep{1.5}{rl}
\Hb_t &= \sum_{j=1}^m(\bar D(z)\vb_{tj})H_j
\\
\Ht_t &= \sum_{j=1}^m\th_{tj} H_j
\\
\th_{tj} &= \Del_j(z)Q(z)\bar D(z)\vb_{tj}
\eea
\eeq
The time-averaged Hamiltonian \refeq{Gavg} is then,
\beq[eq:Gav unc act]
\beasep{1.5}{rl}
\Gb &=
\ds
\sum_{j=1}^m\frac{1}{M}\sum_{t=1}^M
\th_{tj}\Ub_t^\dag H_j\Ub_t
\eea
\eeq
For each $j=1:m$, write the $M\times 1$ vector with elements
$\th_{tj},t=1:M$ as,
\beq[eq:toep th]
\vecc(\th_j) = \Tcal(\Del_j(z))\Tcal(Q(z)\bar D(z))\vecc(\vb)
\eeq
where $\Tcal(P(z))$ is a lower-triangular Toeplitz matrix whose first
column is the impulse response of the LTI system $P(z)$. Since
$\hinfnorm{\Del_j}\leq\del$ implies that
$\normsm{\Tcal(\Del_j(z))}_2\leq\del$ \cite{GrenanderS:58}, the
robustness measure is similar to the form of \refeq{J thmult}, namely,
\beq[eq:J unc act]
\beasep{1.5}{rl}
\Jrbst &= \del\sum_{j=1}^m
\normsm{\Acal_j}_2S_j
\\
\Acal_j &=\mat{\vec A_{1j}&\cdots&\vec A_{Mj}},\quad
A_{tj} = \Ub_t^\dag H_j\Ub_t
\\
S_j &= \normsm{\Kcal\vecc(\vb_j))}_2,\quad
\Kcal = \Tcal(Q(z)\bar D(z))
\eea
\eeq
%

\subsection{Uncertain cross-couplings}

An interesting class of closed systems with multiple errors, which has
similarities to an \emph{open} system, is one where the perturbations
are caused by unwanted cross-coupling interactions amongst parallel
channels. To illustrate this, consider two channels operating at the
same time. With respective dimensions $n_1$ and $n_2$, the system
Hamiltonian is,
\beq[eq:simu pert]
\beasep{1.25}{rcl}
H_t &=& \Hb_t + \Ht_t
\\
\Hb_t &=& \Hb_t^{(1)}\otimes I_{n_2} + I_{n_1}\otimes \Hb_t^{(2)}
\\
\Ht_t &=& \Ht_t^{(1)}\otimes I_{n_2} + I_{n_1}\otimes \Ht_t^{(2)}
+ \Ht_t^{\rm int}
\eea
\eeq
with uncertainty sets,
\beq[eq:simu unc]
\beasep{1.25}{rcl}
\Ht_t^{(i)} &\in& \Hcalunc^{(i)},\ i=1,2
\\
\Ht_t^{\rm int} &\in& \Hcalunc^{\rm int}
\eea
\eeq
The total system nominal unitary is the tensor product of each
unperturbed channel:
$\Ub_t=\Ub_t^{(1)}\otimes\Ub_t^{(2)}$. Consequently, the associated
time-averaged Hamiltonian is,
\beq[eq:Gavg simu]
\beasep{1.5}{rcl}
\Gavg &=& \Gavg^{(1)}\otimes I_{n_2} + I_{n_1}\otimes\Gavg^{(2)}
+ \Gavg^{\rm int}
\\
\Gavg{(i)} &=& \frac{1}{M}\sum_{t=1}^M
(\Ub_t^{(i)})^\dag\Ht_t^{(i)}{\Ub_t^{(i)}},\ i=1,2
\\
\Gavg^{\rm int} &=& \frac{1}{M}\sum_{t=1}^M
  (\Ub_t^{(1)}\otimes\Ub_t^{(2)})^\dag\Ht_t^{\rm int}(\Ub_t^{(1)}\otimes\Ub_t^{(2)})
\eea
\eeq
In general each channel is attempting to make a different unitary. In
this example there are two, denoted by $W_i,i=1,2$. Thus the nominal
fidelity is the product of each as if they are independent,
$\Fnom=\Fnom^{(1)}\Fnom^{(2)}$, or equivalently,
\beq[eq:Fnom1]
\beasep{1.5}{rcl}
\Fnom &=& 
\left|\trace(W^\dag\Ub_N/n_1n_2)\right|^2
\
\left\{
\bea{l}
W = W_1\otimes W_2
\\
\Ub_N = \Ub_N^{(1)}\otimes\Ub_N^{(1)}
\eea
\right.
\eea
\eeq
There is some flexibility in the choice of the robustness
measure. Consider, for example, the worst-case over the
time-averaged Hamiltonians:
\beq[eq:J simu]
\Jrbst = \max\left\{
\norm{\Gavg^{\rm int}},\norm{\Gavg^{(1)}},\norm{\Gavg^{(2)}}
\right\}
\eeq
The norms in the regulation term $\Jrbst$ follow from the specific
character of the uncertainty sets.
This format is easily extended to more channels.

\newpage
\subsection{Open Systems}
\label{sec:open}

The robustness measures for two represntative models of uncertain open
systems are described: bipartite and Lindblad. These are both
presented in the continuous-time framework.
\vspace{-0.25in}
\subsubsection{Bipartite}
\label{sec:bisys}

Consider a bipartite quantum system evolving over the time interval
$t\in[0,T]$ with ``system'' dimension $\ns$ and ``bath' dimension
$\nb$. The continuous-time $n=\ns\nb$ dimensional system-bath unitary
$U(t)$ is obtained from,
\beq[eq:Hopen]
\beasep{1.5}{rl}
i\dot U(t) &= H(t)U(t)
\\
H(t) &= H_{S}(t)\otimes\Ib + \Is\otimes H_{B}(t) + H_{SB}(t)
\eea
\eeq
where $H_S(t)$ is uncertainty-free and control dependent, and where
$H_B(t),H_{SB}(t)$ are uncertain with isolated unitaries,
%
\beq[eq:USUB]
\beasep{1.5}{rl}
i\dot U_S(t) &= H_S(t)U_S(t),
\quad
i\dot U_B(t) = H_B(t)U_B(t)
\eea
\eeq
For a ``system''
unitary target $W_S$, we use the fidelity as defined in
\cite{KosutArenzRabitz:2019},
\beq[eq:Fopen]  
F = \nucnorm{\Gamma/n}^2
\quad
\left\{
\renewcommand{\arraystretch}{1.5}
\bea{l}
\Gamma = \sum_{i=1}^{\ns} V_{[ii]},
\\
V = (W_S\otimes \Ib)^\dag~U(T)
\eea
\right.
\eeq
where $\nucnorm{\cdot}$ is the nuclear-norm (sum of singular values),
and where $\{V_{[ii]},i=1:\ns\}$ are the blosk-diagonal $\nb\times\nb$
sub-matrices of the unitary $V$. (For a closed-system (no bath), this
fidelity reduces to the standard $F=|\tr(W_S^\dag U_S(T)/\ns)|^2$).
In this case it is convenient to define the interaction unitary
as,
\beq[eq:Ropen]
R(t) = (U_S(t)\otimes U_B(t))^\dag U(t)
\eeq
Although in many instances the bath self-dynamics weakly interacts
with the system, \eg, the Hamiltonian $H_B(t)$ is slowly varying over
the operational time, here \refeq{Ropen} is appropriate with $R(t)$
evolving from,
\beq[eq:RGopen]
\beasep{1.5}{rl}
i\dot R(t) &= G(t)R(t)\
\\
G(t) &= (U_S(t)\otimes U_B(t))^\dag H_{SB}(t)(U_S(t)\otimes U_B(t))
\eea
\eeq
Suppose the
time-averaged Hamiltonian $\avg{G}\approx 0$ and the nominal fidelity
$\Fnom=|\tr(W_S^\dag U_S(T)/\ns)|^2\approx 1$. Then $R(T)\approx I$
and $V \approx W_S^\dag U_S(T)\otimes U_B(T) \approx \Is\otimes
U_B(T)$. Under these conditions the fidelity \refeq{Fopen} is
$F\approx 1$.
It follows from these assumptions that the appropriate robustness
measure is,
\beq[eq:Jrbst open]
\Jrbst =
\max_{H_B,H_{SB}\in\Hcalunc}
\norm{\frac{1}{T}\int_0^T G(t)dt}
\eeq
The specific norm depends on the characteristics of $\Hcalunc$.

\subsubsection{Lindblad}
\label{sec:lind unc}

A basic Lindblad master equation describes non-unitary evolution of
an $n\times n$ density matrix $\rho(t),t\in[0,T]$ by,
\beq[eq:lind]
\beasep{1.5}{rl}
\dot\rho &= -i[\Hb(t),\rho] + \Lcal(\rho,\th),\ \rho(0)=\rho_0
\\
\Lcal(\rho,\vec\th) &=
\sum_{k=1}^{K}
\th_k
\left(2L_k\rho L_k^\dag-(L_kL_k^\dag\rho+\rho L_kL_k^\dag)\right)
\eea
\eeq
where $\Hb(t)$ is the nominal (uncertainty-free) Hamiltonian as
defined in \refeq{U}.  Here the Lindblad term $\Lcal(\rho,\vec\th)$ is
an uncertainty model with known operations $L_k\in\Cbf^{n\times
  n},k=1:K$, where $\vec\th$ is a $K\times 1$ vector of uncertain
coefficients $\th_k$ bounded in norm by $\normsm{\vec\th}\leq\del$.

To reconfigure the Lindblad system \refeq{lind} into our standard
multicriterion optimization form \refeq{Jgen}, lift the density to the
vector $x(t)=\vec\rho(t)\in\Cbf^{n^2}$. Then \refeq{lind} becomes,
\beq[eq:xdot]
\beasep{1.5}{rl}
\dot x  &= -iA(t)x + B(\vec\th)x,\ x(0) = \vec\rho_0
\\
A(t) &= I_n\otimes\Hb(t)-\Hb(t)^*\otimes I_n
\\
B(\vec\th) &= \sum_{k=1}^{K}\th_k B_k
\\
B_k &= 
2(L_k^*\otimes L_k)-(I_n\otimes L_k^\dag L_k+(L_k^\dag L_k)^T\otimes I_n)
\eea
\eeq
Note that the nominal (uncertainty-free) system,
\beq[eq:xnom]
\dot\xb = -iA(t)\xb,\ \xb(0) = \vec\rho_0
\eeq
has as solution,
\beq[eq:xunom]
\xb(t) = \Vb(t)\vec\rho_0 \quad
\left\{
\beasep{1.5}{l}
\Vb(t) =  \Ub(t)^*\otimes\Ub(t)
\\
\dot\Ub(t) = -i\Hb(t)\Ub(t),\ \Ub(0)=I_n
\eea
\right.
\eeq
Let the lifted state $x(t) = \Vb(t)R(t)\vec\rho_0$ where $R(t)$, the
lifted $n^2\times n^2$ interaction unitary, evolves from,
\beq[eq:RR]
\beasep{1.5}{rl}
\dot R(t) &= G(t,\vec\th)R(t),\ R(0) = I_{n^2}
\\
G(t,\vec\th) &= \Vb(t)^\dag B(\vec\th) \Vb(t)
= \sum_{k=1}^K\th_k\Vb(t)^\dag B_k \Vb(t)
\eea
\eeq
Here $G(t,\vec\th)$ is the lifted interaction Hamiltonian. A target
unitary $W$ is achieved at the final time $T$ if the density matrix
$\rho(T)\approx W\rho_0W^\dag$ for all $\rho_0$, or in terms of the
lifted state, $x(T)\approx \vecc(W\rho_0W^\dag)=(W^*\otimes
W)\vec\rho_0$. Following \refeq{fidnorm} together with \refeq{RR}
gives the corresponding distance and fidelity measures,
\beq[eq:fidnorm x]
\beasep{1.5}{rcl}
\Ecal &=& \min_{|\phi|=1}
\normsmf{\bar V(T)R(T)-\phi(W^*\otimes W)}^2
\\
&=& 2n^2(1-\Fcal)
\\
\Fcal &=&
\left|
\trace\left((W^*\otimes W)^\dag(\Ub(T)^*\otimes\Ub(T))R(T)\right)/n^2
\right|
\eea
\eeq
Because we have lifted $\rho$ to $\vec\rho$, thereby inflating the
dimension from $n$ to $n^2$, the fidelity here, $\Fcal$, is not squared
as in \refeq{fidnorm}. To see why this is appropriate, if $R(T)\approx
I_{n^2}$, then $\Fcal\approx\Fnom=|\trace(W^\dag\Ub(T)/n)|^2$ as
usual.

Since the uncertainty is bounded, $\normsm{\vec\th}\leq\del$, it
follows that the robustness measure follows from the corresponding
induced matrix norm of the time-averaged interaction Hamiltonian,
%
\beq[eq:JJ]
\Jrbst = \max_{\normsm{\vec\th}\leq\del}\normsm{T\vec{G}_{\rm avg}(\vec\th)}
= \del \norm{\Acal}\quad
\eeq
with
\beq[eq:JJ1]
\beasep{1.5}{l}
\Acal = \mat{\vec\Gam_1&\cdots&\vec\Gam_k}\in\Cbf^{n^4\times K}
\\
\ds
\Gam_k = \int_0^T\Vb(t)^\dag B_k \Vb(t)dt
\eea
\eeq

  \begin{figure*}[t]
  \centering
  \btab{ccc}
  Infidelity \& robustness vs. iterations
  &
  Evaluation of infidelity vs. uncertainty
  &
  Control pulses: nominal vs. optimal
  \\
  \btab{c}
  \includegraphics[width=0.33\textwidth]{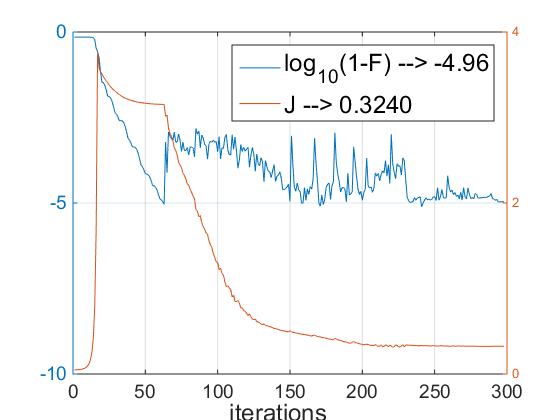}
  \\(a)
  \etab
  &
  \btab{c}
  \includegraphics[width=0.33\textwidth]{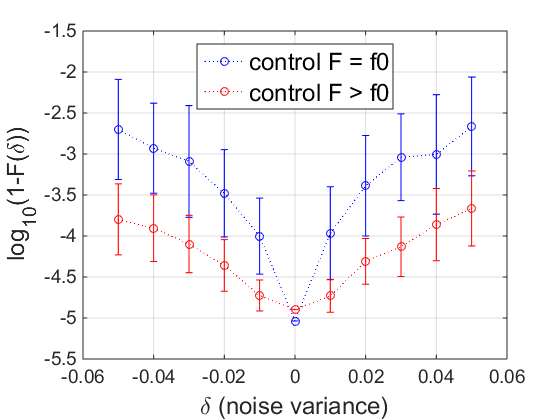}
  \\(b)
  \etab
  &
  \btab{c}
  \includegraphics[width=0.33\textwidth]{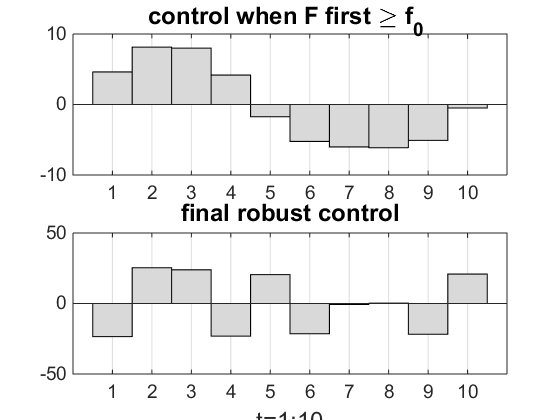}
  \\(c)
  \etab
  \etab
  \caption{{\bf Uncertain time-varying multiplicative control
    parameter} $H_t=(1+\th_t)v_t\sig_x+\sig_z$.}
  \label{fig:fig1tvmult}
  \end{figure*}

  \begin{figure*}[t]
  \centering
  \btab{ccc}
  Infidelity \& robustness vs. iterations
  &
  Evaluation of infidelity vs. uncertainty
  &
  Control pulses: nominal vs. optimal
  \\
  \btab{c}
  \includegraphics[width=0.33\textwidth]{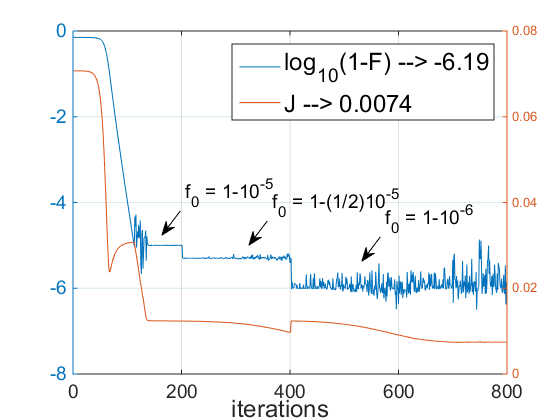}
  \\(a)\etab
  &
  \btab{c}
  \includegraphics[width=0.33\textwidth]{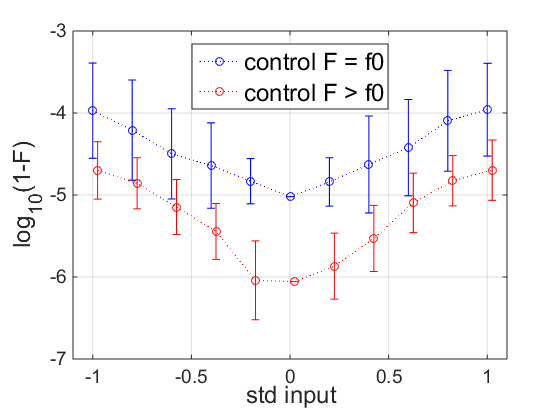}
  \\(b)\etab
  &
  \btab{c}
  \includegraphics[width=0.33\textwidth]{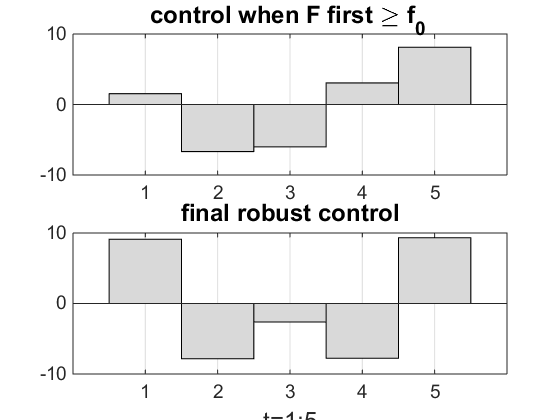}
  \\(c)\etab
  \etab
  \caption{{\bf Two uncertainties} $H_t=
    (1+\th_{x,t})v_t\sig_x+(1+\th_z)\sig_z$.}
  \label{fig:fig1two}
  \end{figure*}

\begin{figure*}[t]
  \centering
  \btab{ccc}
  Infidelity \& robustness vs. iterations
  &
  Evaluation of infidelity vs. uncertainty
  &
  Control pulses: nominal vs. optimal
  \\  
  \btab{c}
  \includegraphics[width=0.33\textwidth]{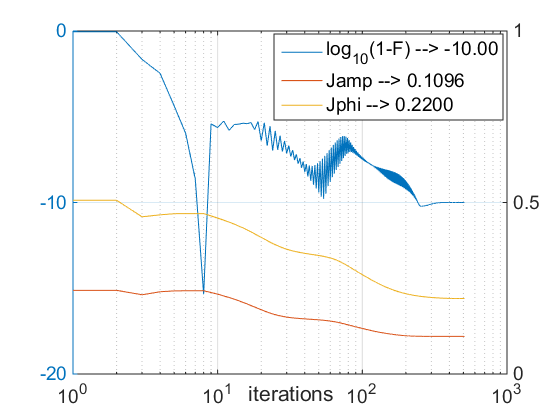}
  \\(a)\etab
  &
  \btab{c}
  \includegraphics[width=0.33\textwidth]{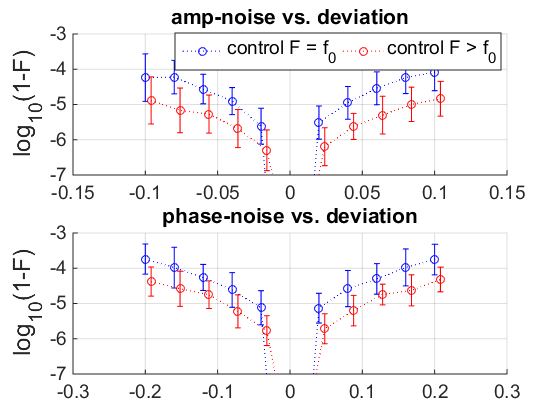}
  \\(b)\etab
  &
  \btab{c}
  \includegraphics[width=0.33\textwidth]{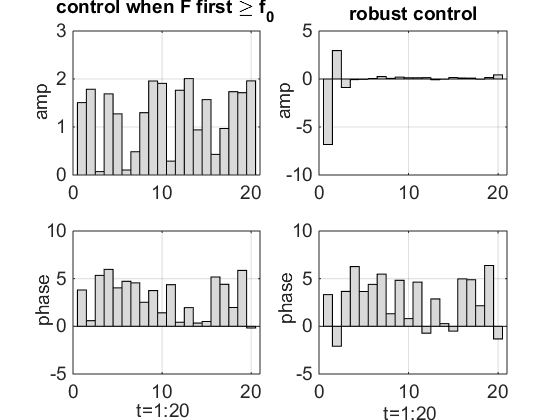}
  \\(c)\etab
  \\
  \btab{c}
  \includegraphics[width=0.33\textwidth]{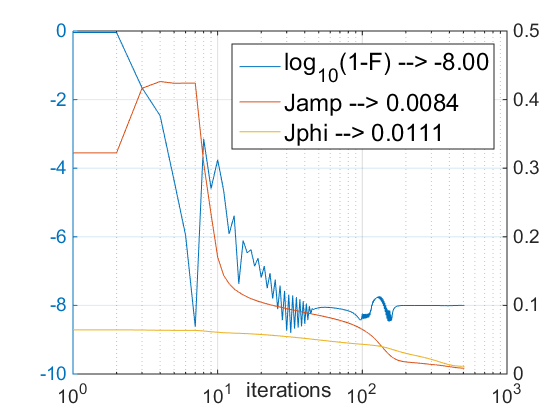}
    \\(d)\etab
    & \btab{c}
    \includegraphics[width=0.33\textwidth]{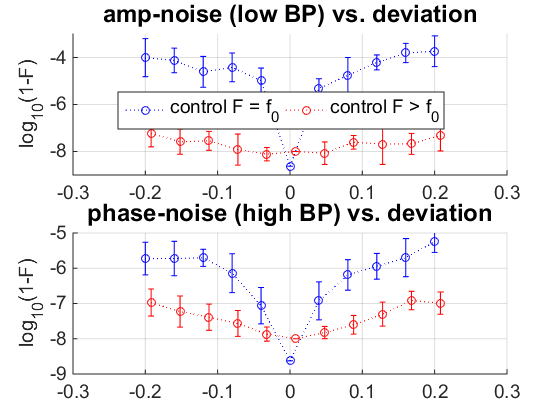}
    \\(e)\etab
    & \btab{c}
    \includegraphics[width=0.33\textwidth]{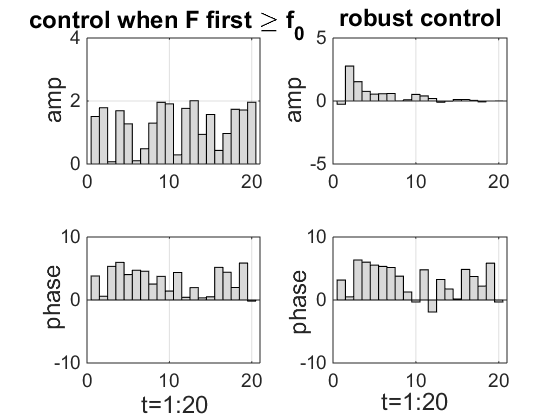}
    \\(f)\etab
    \etab
    \caption{{\bf Two uncertainties.} {\bf(a)-(c) System \refeq{amp
            phi}}: amplitude and phase control are simultaneously
        affected by noise from two \emph{identical} low pass noise
        filters with bandwidth at 1/2 the Nyquist frequency, where
        each filter is driven by independent white noise.  {\bf
          (d)-(f) System \refeq{amp phi}}: amplitude and phase control
        affected simultaneously by two \emph{different} bandpass noise
        filters: low bandpass for amplitude, high bandpass for
        phase. (see \figref{cheby}).}
    \label{fig:twounc2}
\end{figure*}

\section{Numerical examples}
\label{sec:more numex}

\subsection{Multiplicative control noise}

Continuing with the single qubit system, let $\th_t$ denote
multiplicative control noise, that is,
\beq[eq:mult noise]
H_t = (1+\th_t)v_t\sig_x + \sig_z
\eeq
Here we set $N=10$ PWC control pulses, again $M=50$ noise samples, and
the identity target. The multiplicative noise is uniformly random in
$\pm 0.05$ with \emph{five} changes over the time interval as
described by the ``sinc'' spectrum \refeq{Fom ell}. Thus the noise
level randomly changes \emph{twice} during every control pulse
window. The resulting robustness measure is given by \refeq{J thmult}
for $\ell=1:5$ with $\Acal_\ell = (1/M)\sum_{t\in\Tcal_\ell}
v_t\Ub_t^\dag\sig_x\Ub_t$.  For this case of a scalar control, except
for the inclusion of the control pulse magnitude, the robustness
measure reduces to the form \refeq{J thell det}.

\figref{fig1tvmult}(a)-(c) displays the simulation results. These clearly
follow the pattern of \figref{fig1tvmult}(a)-(f) with some notable
differences.  First, the uncertainty improvement at the extremes of
the noise level range $\pm 0.05$ is, on average, one order of
magnitude, unlike the additive noise results which is nearly two
orders of magnitude. Secondly, there is no discernible control pattern
for the robust set of pulses as previously seen in the additive noise
case. These observations may vary from system to system. However, it
is striking that the fidelity error and robustness measure plots per
iteration follow the expected behavior as predicted to be available by
the control landscape topology analysis.


\subsection{Drift and multiplicative uncertainty}

An illustrative single-qubit example of a system with two
uncertainties is,
\beq[eq:Htwounc]
H_t = (1+\th_z)\sig_z + (1+\th_{x,t})v_t\sig_x
\eeq
where $\th_z$ is an uncertain constant parameter (a bias) and
$\th_{x,t}$ is a time-varying multiplicative control noise. The
corresponding uncertainty set is,
\beq[eq:Hcal twounc]
\Hcalunc = \left\{
\Ht_t = \th_z\sig_z + \th_{x,t}v_t\sig_x\
\left|\
\beasep{1.5}{l}
\th_z = \del_z w_z
\\
\vec\th_x = \del_x\Kcal_x\vec w_x
\\
|w_z|\leq 1,\
\norm{w_x}_\infty\leq 1
\eea
\right.
\right.
\eeq
where $\Kcal_x$ is a Toeplitz matrix representation of the noise
dynamics as described in \S\refsec{Hpartv}. Taking the robustness
measure as the sum of the robustness measures for each uncertainty
type yields,
\beq[eq:Jrbst twounc]
\Jrbst=\norm{\Acal S}_\infty\
  \left\{  
  \beasep{1.5}{l}
  \Acal = \mat{\Acal_z&\Acal_x}
  \\
  \Acal_z = \vec A_z,\ A_z = (1/M)\sum_{t=1}^M\Ub_t^\dag\sig_z\Ub_t  
  \\
  \Acal_x = \mat{&\vec A_{x,1}&\cdots&\vec A_{x,M}}
    \\
  A_{x,t} = v_t\Ub_t^\dag\sig_x\Ub_t/M,\ t=1:M
  \\
  S = \mat{\del_z&0\\0&\del_x\Kcal_x}
  \eea
  \right.
  \eeq
  \noindent
  \figref{fig1two}(a)-(c) shows results of the two-stage optimization
  to make the identity gate with $N=5$ PWC control pulses, $M=10$
  averaging samples, a constant drift uncertainty range of
  $\del_z=0.05$, and a multiplicative noise with magnitude
  $\del_x=0.01$ drawn from a uniform distribution with the ``sinc'
  spectrum \refeq{Fom ell} with $L=2$, \ie, two-pulses of width $1/L$.
  Equivalently, the noise dynamics are modeled by the matrix
  $\Kcal_x={\rm blk\_diag}(\un_{M/2},\un_{M/2})$.  Thus $S$ is an
  $M+1\times 3$ matrix.

  As seen in \figref{fig1two}(a), we increased the fidelity threshold
  $f_0$ two times during the iterations which allowed for a continuous
  improvement in robustness.

  \figref{fig1two}(b) shows approximately an order of magnitude
  increase (red), on average, as compared to the noise-free response
  (blue). The data for these plots is from random samples $\th^i=Sw^i,
  i=1,\ldots,100$ with $w^i\in\Rbf^{M+1}$ uniformly random in $\pm
  1$. The final robust control pulses shown in \figref{fig1two}(c)
  reveal an interesting symmetric pattern. This suggests a possible
  deeper connection between this class of uncertainties and
  robustness.

\subsection{Amplitude and phase control noise}
\label{sec:twounc2} 

Consider a single-qubit system having amplitude and phase control
\eg, laser control. A simplified Hamiltonian model is,
\beq[eq:amp phi]
H_t = (\Om_t/2)\left(\sig_x\cos\phi_t+\sig_y\sin\phi_t\right)\
\left\{
\beasep{1.5}{l}
\Om_t=(1+\widetilde\Om_t)\widebar\Om_t
\\
\phi_t = \widebar\phi_t+\widetilde\phi_t
\eea
\right.
\eeq
where $(\widebar\Om_t,\widebar\phi_t)$ are the known control commands
and where $(\widetilde\Om_t,\widetilde\phi_t)$ are time-varying
perturbations, respectively, multiplicative in control and additive in
phase. An approximate Hamiltonian to first-order in these
perturbations is,
\beq[eq:amp phi approx]
H_t \approx \Hb_t + \Ht_t\
\left\{
\beasep{1.5}{l}
\Hb_t = \widebar v_{x,t}\sig_x+\widebar v_{y,t}\sig_y
\\

\Ht_t = \widetilde\Om_t\Hb_t
+ \widetilde\phi_t\Hb_t^\phi
\\
\Hb_t^\phi = \widebar v_{x,t}\sig_y-\widebar v_{y,t}\sig_x
\eea
\right.
\eeq
with $\Hb_t$ the nominal (noise-free) Hamiltonian, a linear function
of the ideal (noise-free) controls $\widebar
v_{x,t}=(\widebar\Om_t/2)\cos\widebar\phi_t$ and $\widebar
v_{y,t}=(\widebar\Om_t/2)\sin\widebar\phi_t$. Use of this approximate
model \refeq{amp phi approx} for control synthesis is not strictly
necessary; it suffices here because the perturbations are small. The
full model \refeq{amp phi} is used for the final analysis and
validation.  Note that both perturbations are effectively in the form
of multiplicative noise as defined in \refeq{Hcal thmult}.

For this example, let both perturbations arise from the output of LTI
filters driven by independent white noise, each drawn from a normal
distribution with respective variances $\del_\Om^2$ and
$\del_\phi^2$. As in \refeq{toep}, the LTI filter dynamics is
contained in the Toeplitz matrices $(\Kcal_\Om,\Kcal_\phi)$ where each
first column is the impulse response. The resulting robustness measure
is of the same form as \refeq{Jrbst twounc}, effectively the sum of
two measures, \ie,
\beq[eq:Jrbst amp phi]
\Jrbst=\normsmf{\Acal S}\
  \left\{  
  \beasep{1.5}{l}
  \Acal = \mat{\Acal_\Om&\Acal_\phi}
  \\
  \Acal_\Om = \mat{\vec A_{\Om,1}&\cdots&\vec A_{\Om,M}}
  \\
  A_{\Om,t} = \Ub_t^\dag\Hb_t\Ub_t/M,\ t=1:M
  \\
  \Acal_\phi = \mat{\vec A_{\phi,1}&\cdots&\vec A_{\phi,M}}
  \\
  A_{\phi,t} = \Ub_t^\dag\Hb_t^\phi\Ub_t/M,\ t=1:M
  \\
  S = \mat{\del_\Om\Kcal_\Om&0\\0&\del_\phi\Kcal_\phi}
  \eea
  \right.
  \eeq

 \figref{twounc2} shows results of the two-stage optimization to make
 the Hadamard gate with $N=20$ PWC control pulses for each of the two
 controls over a time interval of $T=10$ time units. The number of
 averaging samples is $M=200$ and the threshold to switch from Stage-1
 to Stage-2 is set very high at $f_0=1-10^{-10}$.

 In the example shown in \figref{twounc2}(a)-(c) the perturbation
 noise dynamics affecting amplitude and phase are identical low-pass
 filters with bandwidth at 1/2 the Nyquist frequency ($\om_{\rm
   nyq}=\pi/(T/M)$).

\begin{figure}[t]
  \centering
  \includegraphics[width=0.4\textwidth]{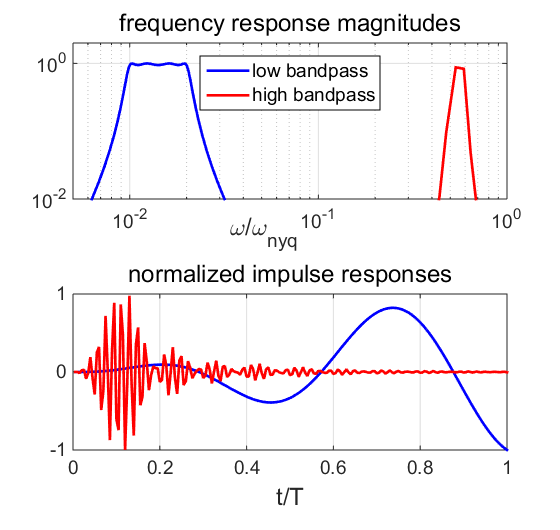}
  \caption{{\bf Chebyshev BandPass Filters} The upper plots shows the
    magnitudes of the low and high BP frequency response
    magnitudes. The lower plot shows the normalized impulse responses:
    slow variations for the low frequency BP and rapid oscillations
    for the high frequency BP.}
  \label{fig:cheby}
\end{figure}

 In the example shown in \figref{twounc2}(d)-(f) the noise filters are
 not identical, both amplitude and phase noise are each independently
 generated from bandpass digital Chebyshev filters whose filter
 frequency magnitudes and (normalized) impulse responses are shown in
 \figref{cheby}. Each filter has 0.5 decibels of peak-to-peak ripple
 in the passband with the amplitude noise passband at the low end of
 the Nyquist frequency, namely $[0.01,0.02]\om_{\rm nyq}$, whereas the
 phase noise passband is at the high end at $[0.5,0.6]\om_{\rm nyq}$.

All the iterations in column 1 of \figref{twounc2} show the (expected)
rapid Stage-1 optimization to make the noise-free system fidelity
reach the selected very high threshold. Then the Stage-2 optimization
turns on whose goal is to hold the threshold while reducing
robustness. The fidelity oscillations occur while the two robustness
measures decrease until an equilibrium or stopping criterion are
reached.

Column 2 of \figref{twounc2} shows the post-optimization evaluation on
the full system. The worst-case fidelities with the bandpass filter
noise are considerably improved in \figref{twounc2}(d)-(f) over those
of the low pass noise analysis seen in \figref{twounc2}(a)-(c). This
behavior is perhaps expected since the noise frequencies are more
concentrated via the narrow frequency ranges of the bandpass noise
filters.

As we have repeatedly remarked, upon viewing the control pulses in
column 3 of \figref{twounc2}, no discernible reason is easily
forthcoming to account for the achieved significant robustness, except
that it was requested of the available null space control resource at
the top of the landscape.

\subsection{Uncertain cross-couplings}
\label{sec:unc cross}

Following \refeq{simu pert}, an example (discrete-time) Hamiltonian of
a two qubit system with uncertain local parameters in each system and
an uncertain cross coupling term is,
\beq[eq:simu ex]
\beasep{1.5}{ll}
H_t = &\left(v_{1t}\sig_x+\th_1\sig_z\right)\otimes I_2
+\  I_2\otimes\left(v_{2t}\sig_x+\th_2\sig_z\right)
\\&
+\ \th_{12}\left(\sig_x^{\otimes 2}+\sig_y^{\otimes 2}+\sig_z^{\otimes 2}\right)
\eea
\eeq
with constant uncertain parameters all in the same range
$(\th_1,\ \th_2,\ \th_{12}) \in [-0.005,0.005]$. The final time is
$T=10$ with $N=20$ PWC pulses per control and $M=40$ uniform samples
for simulation.  The unitary targets for system 1 and 2 are,
respectively, the Hadamard and Identity gates, that is,
$W_1=(\sig_x+\sig_x)/\sqrt{2}$ and $W_2=I_2$. The initial system 1 and
2 controls, $v_{1t}$ and $v_{2t}$, are selected to make these gates
robust locally, \ie, assuming no cross-coupling ($\th_{12}=0$), and
uncertainty only in $(\th_1,\th_2)$. In other words, these initial
controls are robust to the independent variations of $(\th_1,\th_2)$
with no knowledge of cross-coupling uncertainty.

\begin{figure}[h]
  \centering
  \btab{c}
  \includegraphics[width=0.4\textwidth]{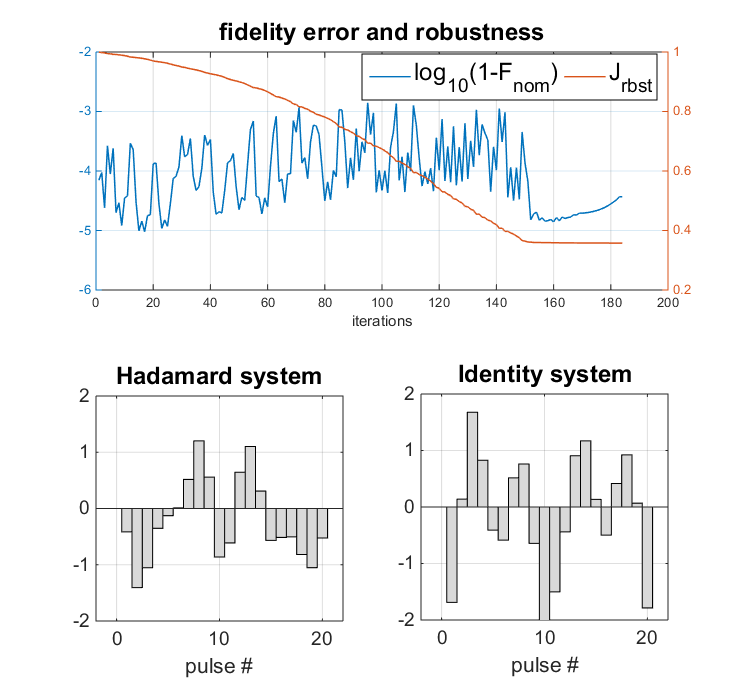}
  \\(a)
  \\
  \includegraphics[width=0.4\textwidth]{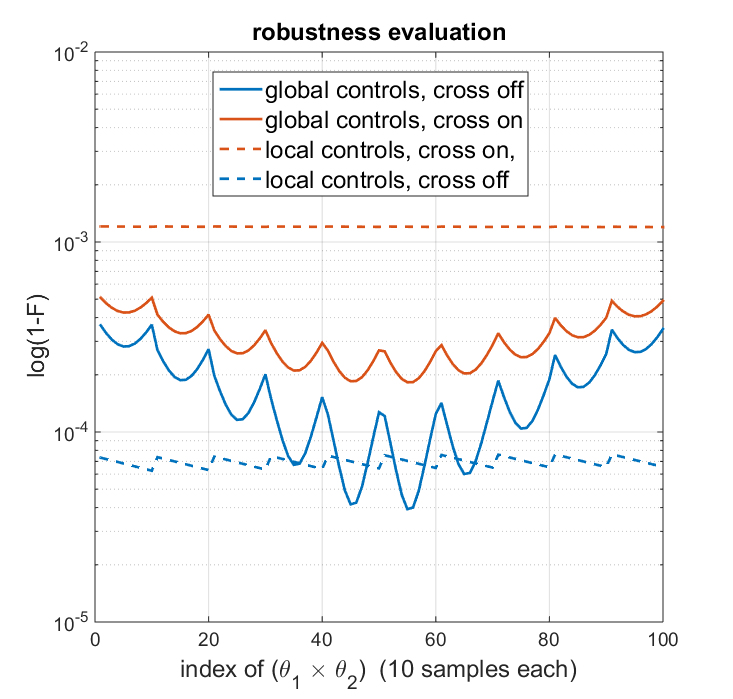}
  \\(b)
  \etab
    \caption{{\bf Cross coupling uncertainty}. {\bf(a)Top.} Fidelity
    error ($1-\Fnom$) and robustness measure ($\Jrbst$)
    vs. iterations. Initial controls are robust to local uncertainty
    only. {\bf(a)Bottom.} Pulse magnitudes of robust control to
    simultaneous local system and cross-coupling uncertainty. {\bf(b)}
    Fidelity error vs. uncertain parameters with independent local
    (dashed) \& global (solid) robust controls.}
  \label{fig:cross}
\end{figure}

The iteration plots in \figref{cross}(a) start with each control being
robust to its own local uncertainty with no knowledge of the
cross-coupling uncertainty. We see again that fidelity remains at a
high fidelity level-set while the time-averaged regulation term
$\Jrbst$ \refeq{J simu} falls until it can no longer be changed.

The results of these individual and independent optimal controls are
shown in \figref{cross}(b) with respect to both the local uncertainty
which is very low (dashed blue plot) whereas considerable degradation
occurs when the cross-coupling is included (dashed red plot) even though it
is a small amount.

By contrast the global robust controls, which account for both local
and cross-coupling uncertainties, show only a slight increase in
fidelity error robustness due to the coupling uncertainty as seen in
\figref{cross}(b) (solid red plots) vs. no coupling (solid blue
plots). With coupling (solid red plot) there is significant robustness
improvement compared to the local robust controls with coupling
present.

 \clearpage

\end{appendix}

\end{document}

%% file: texdefs.tex
\newcommand{\mrm}{\rm}

\newcommand{\mc}{ {m^{}_C} }

\newcommand{\nb}{ {n^{}_B} }

\newcommand{\Rb}{ \bar{R} }

\newcommand{\ns}{{n^{}_S}}

\newcommand{\normf}[1]{\|#1\|_{\mrm fro}}

\newcommand{\Is}{I_S}

\newcommand{\Ib}{I_B}

\newcommand{\Ub}{\bar{U}}

\newcommand{\btab}{\begin{tabular}}
\newcommand{\etab}{\end{tabular}}

\newcommand{\trace}{{\bf Tr}}
\newcommand{\avg}{{\bf E}}
\newcommand{\un}{{\mathbf{1}}}

\newcommand{\tr}{{\mrm tr}~}

\newcommand{\gam}{\gamma}
\newcommand{\alf}{\alpha}
\newcommand{\om}{\omega}
\newcommand{\lam}{\lambda}
\newcommand{\bet}{\beta}
\renewcommand{\th}{\theta}
\newcommand{\del}{\delta}
\newcommand{\sig}{\sigma}

\newcommand{\Del}{\Delta}
\newcommand{\Gam}{\Gamma}
\newcommand{\Om}{\Omega}

\newcommand{\norm}[1]{ \left\| #1 \right\| }
\newcommand{\hinfnorm}[1]{\norm{#1}_{\hinf}}
\newcommand{\hinf}{{\mathbf H}_\infty}

\newcommand{\Vb}{\bar{V}}

\newcommand{\gamb}{\bar{\gam}}

\newcommand{\Ecal}{{\mathcal E}}

\newcommand{\Lcal}{ {\mathcal L} }

\newcommand{\eg}{\emph{e.g.}}
\newcommand{\ie}{\emph{i.e.}}
\newcommand{\etc}{\emph{etc.}}

\newcommand{\bquem}{\begin{quote}\begin{em}}
\newcommand{\equem}{\end{em}\end{quote}}

\newcommand{\blist}{\begin{description}}
\newcommand{\elist}{\end{description}}

\newcommand{\bquote}{\begin{quote}}
\newcommand{\equote}{\end{quote}}

\newcommand{\ben}{\begin{enumerate}}
\newcommand{\een}{\end{enumerate}}

\newcommand{\bit}{\begin{itemize}}
\newcommand{\eit}{\end{itemize}}

\newcommand{\bea}{\begin{array}}
\newcommand{\eea}{\end{array}}

\newcommand{\bds}{\begin{displaystyle}}
\newcommand{\eds}{\end{displaystyle}}

\newcommand{\Rbf}{{\mathbf R}}
\newcommand{\Cbf}{{\mathbf C}}

\newcommand{\ds}{\displaystyle}

\newcommand{\refeq}[1]{(\ref{eq:#1})}

\newcommand{\refsec}[1]{\ref{sec:#1}}

\newcommand{\nom}{{\mrm nom}}

\makeatletter
\def\beq{\@ifnextchar 
[{\@tempswatrue\@beq}{\@tempswafalse\@beq[]}}
\def\@beq[#1]{\begin{equation}\edef\@tmparg{#1}\ifx\@tmparg\@e
mpty \else
	\label{#1}\fi}
\newcommand{\eeq}{\end{equation}}
\makeatother 

\newcommand{\beqaa}{\begin{eqnarray*}}
\newcommand{\eeqaa}{\end{eqnarray*}}

\newcommand{\beqa}{\begin{eqnarray}}
\newcommand{\eeqa}{\end{eqnarray}}

\newcommand{\bc}{\begin{center}}
\newcommand{\ec}{\end{center}}

\newcommand{\bfig}{\begin{figure}}
\newcommand{\efig}{\end{figure}}

\renewcommand{\Is}{I_n}
\renewcommand{\ns}{{n}}
\renewcommand{\nom}{{\rm nom}}

\renewcommand{\normf}[1]{\norm{#1}_{\rm Fro}}



%% file: texdefs1.tex
\makeatletter
\newcommand*\rel@kern[1]{\kern#1\dimexpr\macc@kerna}
\newcommand*\widebar[1]{%
  \begingroup
  \def\mathaccent##1##2{%
    \rel@kern{0.8}%
    \overline{\rel@kern{-0.8}\macc@nucleus\rel@kern{0.2}}%
    \rel@kern{-0.2}%
  }%
  \macc@depth\@ne
  \let\math@bgroup\@empty \let\math@egroup\macc@set@skewchar
  \mathsurround\z@ \frozen@everymath{\mathgroup\macc@group\relax}%
  \macc@set@skewchar\relax
  \let\mathaccentV\macc@nested@a
  \macc@nested@a\relax111{#1}%
  \endgroup
}
\makeatother

\newcommand{\figref}[1]{Figure~\ref{fig:#1}}

\newcommand{\sinc}{{\rm sinc}}

\newcommand{\mat}[1]{
\begin{bmatrix}
#1
\end{bmatrix}
}

\newcommand{\Hb}{\bar{H}}
\newcommand{\Ht}{\tilde{H}}

\newcommand{\xb}{\bar{x}}

\renewcommand{\ns}{{N_S}}
\renewcommand{\nb}{{N_B}}

\newcommand{\Fcal}{{\cal F}}
\newcommand{\Kcal}{{\cal K}}

\newcommand{\Ucal}{{\cal U}}

\newcommand{\Ocal}{{\cal O}}

\renewcommand{\avg}[1]{\langle #1 \rangle}

\renewcommand{\Ub}{{U_B}}

\renewcommand{\Rbf}{\mathbb{R}}
\renewcommand{\Cbf}{\mathbb{C}}

\renewcommand{\trace}{\mbox{\bf Tr}}

\newcommand{\vecc}{\mbox{\bf vec}}

\newtheorem{thm}{Theorem}

\renewcommand{\un}{\mathbbm{1}}

\renewcommand{\th}{\theta}

\newcommand{\Acal}{{\cal A}}
\newcommand{\Tcal}{{\cal T}}